\documentclass[11pt]{article}
\pdfoutput=1

\usepackage{jheppub}
\graphicspath{{./Pictures/}}
\usepackage{subfigure} 

\usepackage{import}

\usepackage{bigints}

\newcommand\xx{\mathrm{x}}
\newcommand\yy{\mathrm{y}}

\title{Perturbative Four-Point Functions from the Analytic Conformal Bootstrap}
\author[1]{Johan Henriksson,}\emailAdd{henriksson@maths.ox.ac.uk}
\author[1]{Tomasz \L ukowski,}\emailAdd{lukowski@maths.ox.ac.uk}

\affiliation[1]{Mathematical Institute, University of Oxford,\\ Andrew Wiles Building, Radcliffe Observatory Quarter,\\ Woodstock Road, Oxford, OX2 6GG, U.K.}

\abstract{We apply the analytic conformal bootstrap method to study weakly coupled conformal gauge theories in four dimensions. We employ twist conformal blocks to find the most general form of the one-loop four-point correlation function of identical scalar operators, without any reference to Feynman calculations. The method relies only on symmetries of the model. In particular, it does not require introducing any regularisation and it is free from the redundancies usually associated with the Feynman approach. By supplementing the general solution with known data for a small number of operators, we recover explicit forms of one-loop correlation functions of four Konishi operators as well as of four half-BPS operators $\mathcal{O}_{\mathbf{20'}}$ in $\mathcal{N}=4$ super Yang-Mills.
}

\begin{document}

\maketitle

\addtocontents{toc}{\protect\setcounter{tocdepth}{1}}

\section{Introduction}

In recent years we have substantially advanced our understanding of conformal field theories (CFT) in dimensions higher than two. Most of the progress comes from the conformal bootstrap approach to CFTs. One successful development is the numerical study of the conformal bootstrap equation \cite{Rattazzi:2008pe} which allowed to find approximate conformal dimensions of a large family of operators, most significantly in the 3d Ising model \cite{ElShowk:2012ht}. Remarkably, analytic methods to solve the bootstrap equation have also been developed recently \cite{Komargodski:2012ek,Fitzpatrick:2012yx,Alday:2013cwa,Alday:2015ota}. They rely on the fact that the large spin sector of a generic CFT is essentially free. This allows to  study the problem as a perturbation theory around infinite spin. An appropriate description is then given by twist conformal blocks \cite{Alday:2016njk}, which resum contributions from all operators with identical classical twist. This reduces the crossing equation to a set of algebraic relations for the CFT data, {\it i.e.~}conformal dimensions and structure constants for all operators in the theory. Twist conformal blocks have already been  successfully used for several theories with slightly broken higher spin symmetry \cite{Alday:2016jfr}, as well as in the large-$N$ expansion of $\mathcal{N}=4$ super Yang-Mills (SYM) \cite{Alday:2017xua}.

In this paper we apply this method to weakly coupled conformal field theories in four space-time dimensions. We study four-point correlation functions
\begin{equation}
\mathcal{G}(x)=(x_{12}^2x_{34}^2)^{\Delta_{\mathcal O}}\langle \mathcal{O}(x_1)\mathcal{O}(x_2)\mathcal{O}(x_3)\mathcal{O}(x_4)\rangle
\end{equation}
 of identical operators built out of fundamental scalar fields of the theory in the small coupling $g$ expansion. Here, $\Delta_{\mathcal{O}}$ is the conformal dimension of the operator $\mathcal{O}$ and $x_{ij}$ denotes the distance between two space-time points. A prototypical example of such theory is $\mathcal{N}=4$ SYM. In order to focus our attention we will discuss two very particular scalar operators in $\mathcal{N}=4$ SYM: the Konishi operator $\mathcal{K}$ and the half-BPS operator $\mathcal{O}_{\mathbf{20'}}$ in the $[0,2,0]$ representation of the $SU(4)$ R-symmetry. Both of them are the simplest gauge invariant scalar operators and have the schematic form $\mathcal{O}=\mbox{Tr}(\phi^2)$, where $\phi$ is a fundamental scalar field of the theory. The methods developed here will however apply to a large class of conformal field theories satisfying a set of assumptions spelled out at the end of this section.

In the following we study four-point correlation functions in the perturbation theory around vanishing coupling constant $g=0$,
\begin{equation}
\mathcal{G}(x)=\mathcal{G}^{(0)}(x)+g\, \mathcal{G}^{(1)}(x)+\ldots.
\end{equation}
 The leading-order answers $\mathcal{G}^{(0)}(x)$ can be found by directly performing Wick contractions and depend on a single parameter related to the central charge of the theory. In this paper we focus most of our attention on the one-loop function $\mathcal{G}^{(1)}(x)$ and find its general form using only the conformal symmetry, crossing symmetry and the structure of the operator product expansion (OPE). In the two cases that we study we find a family of crossing-symmetric solutions which depend on a small number of free parameters. The most transcendental part of the answer is given by the so-called box function times a rational function. These has to be supplemented by lower transcendental functions. We find the explicit form of these functions without referring to Feynman diagram calculations. In particular, we will avoid introducing any regularisation or any redundancies fundamentally bound to the Feynman approach. In order to find a particular four-point correlator we supplement our general solution with a few explicit values of the CFT data for operators with small classical conformal dimension and spin. These can be found in the literature.

Our method will be based on only a few assumptions and will use properties of conformal field theories, which we summarise here:
\begin{itemize}
\item We study unitary weakly coupled conformal gauge theories in four dimensions.
\item We use the fact that four-point correlation functions are crossing symmetric.
\item We use the knowledge of the OPE structure. Furthermore, we rely on an explicit form of the conformal blocks in four dimensions and the superconformal blocks for the half-BPS operators $\mathcal{O}_{\mathbf{20'}}$ in $\mathcal{N}=4$ SYM.
\item We assume that infinite towers of operators parametrised by spin $\ell$ have a regular expansion of the CFT data at large spin, {\it i.e.~}the CFT data can be written as a Taylor expansion of $\frac{1}{\ell}$ with possible $\log \ell$ insertions. 
\end{itemize}
It was already found in \cite{Heemskerk:2009pn,Alday:2014tsa} that there exists a class of crossing symmetric solutions which correspond to CFT data that is truncated in spin. In particular, the instanton solutions are of this type, as shown in \cite{Arutyunov:2000im}. Our analysis extends these results by including also solutions unbounded in spin. Since crossing at one loop in perturbation theory is a linear problem, we can treat these two types of solutions separately and focus only on the latter.

The paper is organised as follows: in section \ref{sec:four.points} we collect basic information about four-point correlation functions and their properties. In section \ref{sec:twist.conformal.blocks} we introduce the notion of twist conformal blocks and H-functions and study their properties. In section \ref{sec:FindingNemo} we use H-functions to find a family of solutions to the conformal bootstrap equation and in particular recover the known form of the four-point correlator of Konishi operators. In section \ref{sec:super.case} we repeat the analysis from the previous two sections in the case of the correlation function of four half-BPS operators $\mathcal{O}_{\mathbf{20'}}$ in $\mathcal{N}=4$ SYM. We end the paper with conclusions and outlook and supplement it with a few appendices containing the more technical ingredients of our results.

\section{Four-Point Correlators}
\label{sec:four.points}

In this section we collect all relevant information about four-point correlation functions of operators that we will study in the rest of this paper. In the first part we describe four-point correlators of four identical scalar operators with classical dimension $\Delta_0=2$. This is relevant for the Konishi operator in $\mathcal{N}=4$ SYM, which is of the form 
\begin{equation}
\mathcal{K}(x)=\mbox{Tr}(\phi_I(x) \phi^I(x)),
\end{equation}
where $I$ is the $SO(6)$ R-symmetry index. We study the correlation function of four Konishi operators using the ordinary conformal partial wave decomposition in four dimensions \cite{Dolan:2000ut}. 

In the second part we study the $\mathcal{N}=4$ SYM half-BPS operator in the $[0,2,0]=\mathbf{20'}$ representation of the $SU(4)$ R-symmetry
\begin{equation}
\mathcal{O}_{\mathbf{20'}}(x,y)=y_I\,y_J\,\mbox{Tr}(\phi^I(x)\phi^J(x))\,,
\end{equation}
where we have introduced an auxiliary six-dimensional complex null vector $y_I$, namely $y\cdot y\equiv y_Iy^I=0$. In order to properly accommodate for a non-trivial R-symmetry structure of the correlation function of four half-BPS operators we employ superconformal blocks introduced in \cite{Dolan:2004iy}.

\subsection{Conformal Partial Wave Decomposition for Konishi Operators}
\label{sec:bosonic.four.points}
First, let us consider the case relevant for the Konishi operator $\mathcal{K}$, namely a scalar operator with the conformal dimension 
\begin{equation}
\Delta_\mathcal{K}=2+\sum_{i=1}^\infty \gamma^{(i)}_{\mathcal{K}}\, g^i\,.
\end{equation}
 From conformal invariance the four-point correlator of identical scalar operators takes the form
\begin{equation}\label{four.point.function}
\langle \mathcal{K}(x_1)\mathcal{K}(x_2)\mathcal{K}(x_3)\mathcal{K}(x_4)\rangle=\frac{\mathcal{G}(u,v)}{x_{12}^{2\Delta_{\mathcal{K}}}x_{34}^{2\Delta_{\mathcal{K}}}},
\end{equation}   
where the cross-ratios $u$ and $v$ are defined by\footnote{In this paper we use the symbols $\xx_1$ and $\xx_2$ instead of the more standard notation $z$ and $\bar z$.}
\begin{equation}\label{eq.crossratios}
u=\xx_1 \,\xx_2=\frac{x_{12}^2x_{34}^2}{x_{13}^2x_{24}^2},\qquad v=(1-\xx_1)(1-\xx_2)=\frac{x_{14}^2x_{23}^2}{x_{13}^2x_{24}^2}\,.
\end{equation}
In the following we will use both sets of cross-ratios $(u,v)$ and $(\xx_1,\xx_2)$ interchangeably.
Crossing symmetry demands that the four-point function \eqref{four.point.function} is invariant under exchange of positions of any two operators. Since we study four identical operators, it leads to two independent conditions satisfied by the correlation function \eqref{four.point.function}:
\begin{equation}\label{crossing.symmetry}
\mathcal{G}(u,v)=\mathcal{G}\left(\frac{u}{v},\frac{1}{v}\right)\,,\qquad v^{\Delta_{\mathcal{K}}}\mathcal{G}(u,v)=u^{\Delta_{\mathcal{K}}}\mathcal{G}(v,u)\,.
\end{equation}
In the following, we will solve these equations and study their solutions as perturbations around small $u\sim 0$ and $v\sim 0$, corresponding to $\xx_1\sim0$ and $\xx_2\sim1$. While the first equation in \eqref{crossing.symmetry} can easily be expanded using the conformal partial wave decomposition, the second equation has to be treated more carefully. In order to do that we will need to employ the twist conformal blocks introduced in \cite{Alday:2016njk}. We refer to the second equation in \eqref{crossing.symmetry} as the {\it conformal bootstrap equation}. 

The conformally invariant function $\mathcal{G}(u,v)$ entering \eqref{four.point.function} admits a decomposition into conformal partial waves obtained by considering the OPE expansion in the limit $x_1\to x_2$
\begin{equation}\label{eq:BlockDecomposition}
\mathcal G(u,v)=\sum_{\tau,\ell,i}a_{\tau,\ell,i}\,G_{\tau,\ell}(u,v).
\end{equation}
Here the sum runs over all conformal primaries of twist $\tau=\Delta-\ell$, where $\Delta$ is the conformal dimension, and even spin $\ell$ present in the OPE decomposition of two Konishi operators 
\begin{equation}
\mathcal{K}\times \mathcal{K}\sim\sum_{\tau,\ell,i}C_{\mathcal{K}\mathcal{K}\mathcal{O}_{\tau,\ell,i}}\left(\mathcal{O}_{\tau,\ell,i}+\ldots\right),
\end{equation}
where the $\ldots$ stands for contributions from descendants of $\mathcal{O}_{\tau,\ell,i}$. The index $i=1,\ldots,d_{\tau_0,\ell}$ runs over a possible additional degeneracy in the spectrum of operators with a given twist and spin.
We denoted the square of OPE coefficients by $a_{\tau,\ell,i}=C^2_{\mathcal{K}\mathcal{K}\mathcal{O}_{\tau,\ell,i}}$. 
The conformal blocks $G_{\tau,\ell}(u,v)$, which resum contributions coming from all descendants of a given conformal primary operator, can be found explicitly for four dimensions \cite{Dolan:2000ut}. For even spins they take the following form
\begin{equation}\label{eq:ConformalBlock}
G_{\tau,\ell}(\xx_1,\xx_2)=\frac{\xx_1 \xx_2}{\xx_1-\xx_2}\left(k_{\frac\tau2+\ell}(\xx_1)k_{\frac \tau2-1}(\xx_2)-k_{\frac\tau2+\ell}(\xx_2)k_{\frac \tau2-1}(\xx_1)\right)\,,
\end{equation}
where $k_\beta(\xx)=\xx^\beta{_2F_1}(\beta,\beta,2\beta;\xx)$ and ${_2F_1}(a,b,c;\xx)$ is a hypergeometric function. 
It is easy to check that each conformal block satisfies the first equation in \eqref{crossing.symmetry}.

\begin{figure}[t!]
\centering
\includegraphics{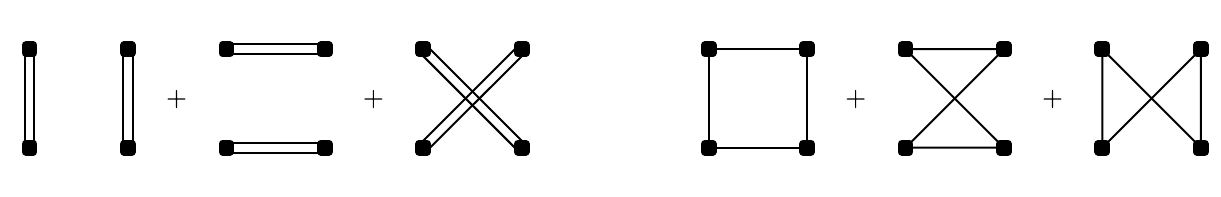}
\caption{Wick contractions relevant for the tree-level calculation.}
\label{fig:Born}
\end{figure}

On the other hand, in perturbative conformal gauge theories the four-point function \eqref{four.point.function} admits a small coupling expansion
\begin{equation}\label{four.point.weak}
\mathcal{G}(u,v)=\mathcal{G}^{(0)}(u,v)+g \,\mathcal{G}^{(1)}(u,v)+\ldots,
\end{equation}
where $g$ is the gauge coupling.
The tree level term can be directly evaluated using Wick contractions in the free theory as in Fig.~\ref{fig:Born}
\noindent and renders
\begin{equation}\label{tree.level}
\mathcal G^{(0)}(u,v)=\left(1+u^2+\frac{u^2}{v^2}\right)+c\left(u+\frac uv+\frac{u^2}v\right),
\end{equation}
where $c$ is a theory-dependent constant which for example for $\mathcal{N}=4$ SYM with gauge group $SU(N)$ is proportional to the inverse of the central charge, 
$c\sim (N^2-1)^{-1}$.
Performing the conformal partial wave decomposition we find that for each classical twist $\tau_0=2,4,6,\ldots$ there exists an infinite tower of operators contributing to the sum in \eqref{eq:BlockDecomposition}, labelled by spin $\ell$ and degeneracy index $i$. This twist degeneracy will be partially lifted in the next sections, when we include perturbative corrections to the four-point correlator. Using the conformal partial wave decomposition of \eqref{tree.level} we can compute the tree-level structure constants. They are non-zero only for even spins $\ell$ and take the form
\begin{align}\label{eq:structureconstants}
\langle a^{(0)}_{\tau_0,\ell}\rangle=\begin{cases} 2c\,\frac{\Gamma(\ell+\frac{\tau_0}{2})^2}{\Gamma(2\ell+\tau_0-1)}\,,&\tau_0=2\,, \vspace{0.2cm}
\\
2\frac{\Gamma(\frac{\tau_0}{2}-1)^2\Gamma(\frac{\tau_0}{2}+\ell)^2}{\Gamma(\tau_0-3)\Gamma(\tau_0+2\ell-1)}\left(c\,(-1)^{\frac{\tau_0}{2}}+(\tau_0+\ell-2)(\ell+1)\right)\,,& \tau_0>2\,,\end{cases}
\end{align}
where we have introduced an average of structure constants over operators with the same classical twist and spin, $\langle a^{(0)}_{\tau_0,\ell}\rangle\equiv\sum_i a^{(0)}_{\tau_0,\ell,i}$. Notice that from the correlator \eqref{tree.level} alone it is not possible to calculate individual structure constants by this procedure. 

In the following sections we will find the most general one-loop correction to \eqref{tree.level} using the conformal symmetry, crossing symmetry and the structure of the OPE. In particular, we will compute an explicit form of the perturbative corrections to the structure constants $\langle a^{(0)}_{\tau_0,\ell}\rangle \to\langle a^{(0)}_{\tau_0,\ell}\rangle+g\langle a^{(1)}_{\tau_0,\ell}\rangle$ as well as to the twists $\tau_0\to\tau_0+g \,\frac{\langle a^{(0)}_{\tau_0,\ell} \gamma^{(1)}_{\tau_0,\ell}\rangle}{\langle a^{(0)}_{\tau_0,\ell}\rangle}$. The knowledge of results for individual operators $\mathcal{O}_{\tau,\ell,i}$ will not be necessary to find the complete four-point correlator at one loop, they will become relevant only at the two-loop order. We will comment on this matter in the outlook of this paper.

\subsection{Superconformal Partial Wave Decomposition for Half-BPS Operators}
\label{sec:super.four.points}
As the second example, we consider the four-point correlation function of four half-BPS operators $\mathcal{O}_{\mathbf{20'}}$ in $\mathcal{N}=4$ SYM, which are protected and their dimension is $\Delta_{\mathcal{O}_{\mathbf{20'}}}=2$. The four-point correlation function of such operators decompose into the following two contributions
\begin{equation}\label{eq:super.four.point}
\langle \mathcal{O}_{\mathbf{20'}}(x_1,y_1)\mathcal{O}_{\mathbf{20'}}(x_2,y_2)\mathcal{O}_{\mathbf{20'}}(x_3,y_3)\mathcal{O}_{\mathbf{20'}}(x_4,y_4)\rangle=\mathcal{G}_{\mathrm{Born}}(x,y)+\mathcal{G}_{\mathrm{pert}}(x,y),
\end{equation}
where $\mathcal{G}_{\mathrm{pert}}(x,y)$ vanishes when $g\to0$. The part $\mathcal{G}_{\mathrm{Born}}(x,y)$ corresponds to the Born approximation and is a rational function of space time and R-symmetry coordinates. Again, it can be evaluated directly by Wick contractions and it boils down to the same set of graphs as in Fig.~\ref{fig:Born}. It renders
\begin{equation}
\mathcal{G}_{\mathrm{Born}}(x,y)=d_{12}^2d_{34}^2+d_{13}^2d_{24}^2+d_{14}^2d_{23}^2+\tilde c\,\big(d_{12}d_{23}d_{34}d_{14}+d_{12}d_{24}d_{34}d_{13}+d_{13}d_{24}d_{23}d_{14}\big),
\end{equation}
where the superpropagator $d_{ij}$ is given by
\begin{equation}
d_{ij}=\frac{y_{ij}^2}{x_{ij}^2},\qquad y_{ij}=y_i\cdot y_j,
\end{equation}
and $\tilde c$ is a theory-dependent constant which for $SU(N)$ $\mathcal N=4$ SYM again depends only on the central charge $\tilde c\sim (N^2-1)^{-1}$.

From the superconformal Ward identities \cite{Nirschl:2004pa}, the interacting part of the four-point function can be written in a factorised form 
\begin{equation}
\mathcal{G}_\mathrm{pert}(x,y)=d_{12}^2d_{34}^2 \frac{(\xx_1-\yy_1)(\xx_1-\yy_2)(\xx_2-\yy_1)(\xx_2-\yy_2)}{(\yy_1\, \yy_2)^2}\mathcal{H}(u,v)\,,
\end{equation}
where we have introduced a set of cross-ratios for the R-symmetry coordinates
\begin{equation}
\yy_1 \yy_2=\frac{y_{12}^2y_{34}^2}{y_{13}^2y_{24}^2},\qquad (1-\yy_1)(1-\yy_2)=\frac{y_{14}^2y_{23}^2}{y_{13}^2y_{24}^2}\,.
\end{equation}
Similar to the four-point function of Konishi operators, crossing symmetry implies that the function $\mathcal{H}(u,v)$ satisfies the two equations
\begin{equation}\label{eq:superbootstrap}
 \mathcal{H}(u,v)=\frac{1}{v^2}\mathcal{H}\left(\frac{u}{v},\frac{1}{v}\right)\,,\qquad v^{2}\mathcal{H}(u,v)=u^{2}\mathcal{H}(v,u)\,,
\end{equation} 
where in the second equation we used explicitly the fact that $\Delta_{\mathcal{O}_{\mathbf{20'}}}=2$.

On the other hand, the four-point correlation function~\eqref{eq:super.four.point} admits a superconformal partial wave decomposition, see e.g. \cite{Doobary:2015gia} 
\begin{equation}
\mathcal{G}_{\mathrm{Born}}(x,y)+\mathcal{G}_{\mathrm{pert}}(x,y)=d_{12}^2d_{34}^2 \sum_{\mathcal{R},i}A_{\mathcal{R},i}\,\mathcal{S}_{\mathcal{R}}(x,y),
\end{equation}
where the sum runs over all superconformal primary operators appearing in the OPE expansion of two half-BPS operators
\begin{equation}\label{eq:super.OPE}
\mathcal{O}_{\mathbf{20'}}\times \mathcal{O}_{\mathbf{20'}}\sim\sum_{\mathcal{R},i}C_{\mathcal{O}_{\mathbf{20'}}\mathcal{O}_{\mathbf{20'}}\mathcal{O}_{\mathcal{R},i}}\left(\mathcal{O}_{\mathcal{R},i}+\ldots\right).
\end{equation}
Superconformal primaries in \eqref{eq:super.OPE} are labelled by their twist $\tau=\Delta-\ell$, spin $\ell$ and a representation of the $SU(4)$ R-symmetry of $\mathcal{N}=4$ SYM, which we collectively denote by $\mathcal{R}$. Again, we also introduced the label $i$  which takes care of a possible additional degeneracy of operators with the same twist, spin and the R-symmetry label. Importantly, the superconformal blocks do not depend on the label $i$. An explicit description of superconformal multiplets and an explicit form of the superconformal blocks $\mathcal{S}_{\mathcal{R}}$ can be found in the appendix~\ref{app:superblocks}. As it is summarised there, we distinguish three types of supermultiplets in \eqref{eq:super.OPE}: half-BPS, quarter-BPS and long supermultiplets. All half-BPS and most quarter-BPS supermultiplets have their conformal dimensions and structure constants protected by supersymmetry. Then, their two-point and three-point correlation functions are completely determined by the Born approximation $\mathcal{G}_\mathrm{Born}$. They will therefore not contribute to the interacting  part $\mathcal{H}(u,v)$ of the four-point correlation function. The only exception are quarter-BPS supermultiplets at the unitarity bound. They can combine in the interacting theory to form a long, non-protected supermultiplet \cite{Dolan:2002zh,Heslop:2003xu}. This is exactly the case for the twist-two operators. Together with the other long supermultiplets they form a complete non-protected spectrum of operators present in the intermediate channel. Since we want to find the one-loop correction to $\mathcal{H}(u,v)$, we will in the following be interested only in the non-protected part of the spectrum. 

We can perform a superconformal partial wave decomposition of the leading contribution $\mathcal{G}_{\mathrm{Born}}(x,y)$ to the four-point function and get structure constants for all non-protected multiplets
\begin{equation}\label{eq:structure.constants.born}
 \langle A_{\tau_0,\ell}^{(0)}\rangle=\begin{cases} 2\tilde c\,\frac{\Gamma(\ell+\frac{\tau_0}{2}+2)^2}{\Gamma(2\ell+\tau_0+3)}\,,&\tau_0=2\,,
\\2\frac{\Gamma(\frac{\tau_0}{2}+1)^2\Gamma(\frac{\tau_0}{2}+\ell+2)^2}{\Gamma(\tau_0+1)\Gamma(\tau_0+2\ell+3)}\left(\tilde c\,(-1)^{\frac{\tau_0}{2}}+(\tau_0+\ell+2)(\ell+1)\right),& \tau_0=4,6,8,\ldots\,.\end{cases}
\end{equation}
It is interesting to notice that $\langle A^{(0)}_{\tau_0,\ell}\rangle=\langle a^{(0)}_{\tau_0,\ell}\rangle\big|_{c\to\tilde c,\tau_0\to\tau_0+4}$.

Furthermore, using the explicit form of superconformal blocks \eqref{eq:super.long} and \eqref{eq:twist2.recomb} for non-protected multiplets, the interacting part of the four-point correlation function can be expanded as
\begin{equation}\label{eq:superconformal.decomposition}
\mathcal H(u,v)=\sum_{\tau,\ell}\langle A_{\tau,\ell}\rangle u^{-2} \,G_{\tau+4,\ell}(\xx_1,\xx_2),
\end{equation}
where $G_{\tau,\ell}(\xx_1,\xx_2)$ is exactly the same conformal block as in \eqref{eq:ConformalBlock} in section \ref{sec:bosonic.four.points}. We notice that both leading-order structure constants $\langle A^{(0)}_{\tau_0,\ell}\rangle$ and superconformal blocks for non-protected supermultiplets are related to the Konishi case by shifting $\tau_0\to\tau_0+4$. For this reason, the one-loop calculation for the four-point correlator of half-BPS operators is analogous to a similar analysis for four Konishi operators, after this shift is implemented at the level of twist conformal blocks. 

\section{Twist Conformal Blocks}
\label{sec:twist.conformal.blocks}
In this section we describe twist conformal blocks and their generalisations introduced in \cite{Alday:2016njk} and use them to rewrite the conformal partial wave decomposition of four-point correlation functions from the previous section. We focus in this section exclusively on the case of four Konishi operators, leaving the half-BPS case to section~\ref{sec:super.case}. We start by defining twist conformal blocks relevant for the tree-level correlators and then define their generalisations with spin-dependent insertions that will be relevant for the perturbative expansion around the tree-level solution. 

\subsection{Twist Conformal Blocks}
A motivation to study twist conformal blocks is the observation that in perturbation theory there exists, for each even number $\tau_0=2,4,6,\ldots$, an infinite family of operators $\mathcal{O}_{\tau_0,\ell,i}$, $\ell=0,2,4,\ldots$, $i=1,\ldots,d_{\tau_0,\ell}$, with the classical twist equal to $\tau_0$:
\begin{equation}
\tau=\tau_0+\mathcal{O}(g)\,.
\end{equation}
Therefore, at tree-level we have an infinite twist degeneracy which is lifted only when we turn on the coupling constant. In particular, it motivates us to resum contributions coming from all intermediate operators with the same classical twist $\tau_0$. In this case, the leading order four-point correlator \eqref{tree.level} can be decomposed as
\begin{equation}
\mathcal G^{(0)}(u,v)=\!\sum_{\tau_0=2,4,\ldots}\! H_{\tau_0}(u,v),
\end{equation}
where we have defined {\it twist conformal blocks} 
\begin{equation}\label{eq:Hfun.def}
H_{\tau_0}(u,v)=\sum_{\ell=0}^\infty \langle a^{(0)}_{\tau_0,\ell}\rangle G_{\tau_0,\ell}(u,v),
\end{equation}
with $\langle a^{(0)}_{\tau_0,\ell}\rangle$  given in \eqref{eq:structureconstants}. The sum in \eqref{eq:Hfun.def} can be  performed for any $\tau_0$ using the explicit form of conformal blocks. For example for $\tau_0=2$ it renders
\begin{equation}\label{eq.H2expl}
H_2(u,v)=c\frac{u}{v}+c\, u .
\end{equation}
For higher twists the explicit form of $H_{\tau_0}(u,v)$ is more involved and we will not present it here. However, in all subsequent calculations we will need only their power divergent part as $v\to 0$. Such divergent parts can be easily calculated and written in a closed form as we will show below.

\subsection{H-functions}
In order to study perturbative corrections to the tree-level correlation function $\mathcal{G}^{(0)}(u,v)$ we need to generalise the notion of twist conformal blocks. In particular, when the coupling constant $g$ is not zero, the twist degeneracy we observed at the tree level is lifted and each $\mathcal O_{\tau_0,\ell,i}$ gets individual corrections to their twists and structure constants,
\begin{align}
\tau_{\tau_0,\ell,i}&=\tau_0+g\,\gamma^{(1)}_{\tau_0,\ell,i}+\mathcal O(g^2),
\\
a_{\tau_0,\ell,i}&=a^{(0)}_{\tau_0,\ell,i}+g \,a^{(1)}_{\tau_0,\ell,i}+\mathcal O(g^2).
\end{align}
Here $\gamma^{(1)}_{\tau_0,\ell,i}$ is the one-loop anomalous dimension of $\mathcal O_{\tau_0,\ell,i}$ and $a^{(1)}_{\tau_0,\ell,i}$ is the one-loop correction to the structure constants. In the conformal partial wave decomposition, these corrections will introduce an additional dependence on the spin and will modify the sum in the definition of the twist conformal blocks. Therefore, we will need to calculate sums of the form
\begin{equation}\label{eq:sum.with.insertions}
\sum_{\ell=0}^\infty \langle a^{(0)}_{\tau_0,\ell}\rangle  \kappa_{\tau_0}(\ell)\,G_{\tau_0,\ell}(u,v) ,
\end{equation}
where $\kappa_{\tau_0}(\ell)$ stands for the spin dependence coming from either the anomalous dimensions or the OPE coefficients. 
In particular, these insertions can be of two kinds: unbounded in spin $\ell$ or truncated contributions with finite support in $\ell$. The truncated contributions do not affect the divergent part of correlator and we will postpone their study to the following section. On the other hand, for the insertions unbounded  in spin the sum \eqref{eq:sum.with.insertions} can be calculated as an expansion around the infinite value of spin. In particular, in the unbounded case $\kappa_{\tau_0}(\ell)$ can be expanded around large values of the  eigenvalue  $J_{\tau_0}^2=(\frac{\tau_0}{2}+\ell)(\frac{\tau_0}{2}+\ell-1)$ of a shifted quadratic Casimir of the conformal group:
\begin{equation}
\kappa_{\tau_0}(\ell)=\sum_{m=0}^\infty \left( \frac{C_{(m)}}{J_{\tau_0}^{2m}}+\frac{C_{(m,\log)}}{J_{\tau_0}^{2m}}\log J_{\tau_0}+\ldots\right),
\end{equation}
as was shown in \cite{Alday:2015eya}.
Then, in order to study perturbation theory beyond the tree level, we consider a set of functions \cite{Alday:2016njk}
\begin{equation}\label{eq:HfunctionsDef}
H^{(m,\log^n)}_{\tau_0}(u,v)=\sum_\ell \langle a_{\tau_0,l}^{(0)}\rangle \frac{(\log J_{\tau_0})^n}{J_{\tau_0}^{2m}}G_{\tau_0,l}(u,v),
\end{equation}
which we will refer to as {\em H-functions}. The H-functions describe contributions from an infinite sum of conformal blocks with spin-dependent insertions. In the case $m=n=0$ the H-functions $H_{\tau_0}^{(0)}(u,v)$ coincide with the twist conformal blocks. Importantly, the functions \eqref{eq:HfunctionsDef} satisfy the following recursion relation 
\begin{equation}\label{eq:Hrec}
H_{\tau_0}^{(m,\log^n)}(u,v)=\mathcal C H_{\tau_0}^{(m+1,\log^n)}(u,v),
\end{equation}
where we defined the shifted quadratic Casimir
\begin{equation}\label{eq:fullCasimir}
\mathcal C=D_1+D_2+2\frac{\xx_1 \xx_2}{\xx_1-\xx_2}\left((1-\xx_1)\partial_1-(1-\xx_2)\partial_2\right)-\frac{\tau_0(\tau_0-6)}4,
\end{equation}
with $D_i=(1-\xx_i)\xx_i^2\partial^2_i-\xx_i^2\partial_i$ and $(\xx_1,\xx_2)$ are defined in \eqref{eq.crossratios}.  The relation \eqref{eq:Hrec} can be easily proven by noticing that each individual conformal block $G_{\tau_0,\ell}(\xx_1,\xx_2)$ is an eigenvector of the Casimir operator $\mathcal{C}$ with the eigenvalue $J_{\tau_0}^2$.

\subsection{Enhanced Divergences}\label{sec:enhanced.div}
In the following we will not need an explicit form of the functions $H^{(m,\log^n)}_{\tau_0}(u,v)$ but only their enhanced divergent part as $v\to0$. Expanding \eqref{eq:ConformalBlock} in this limit, one can notice that the conformal blocks behave as a logarithm $G_{\tau_0,\ell}(u,v)\sim \log(v)$ for $v\to0$. By enhanced divergence we will mean terms which cannot be written as a finite sum of conformal blocks. There are two kinds of enhanced divergences we will encounter: inverse powers of $v$, and functions with higher powers of the logarithm, that is functions of the form $p(v)\log^n v$, $n>1$, where $p(v)$ is regular for $v\to0$. 
As was shown in \cite{Alday:2015eya}, the power divergent part of $H_{\tau_0}(u,v)$ is completely determined by operators with large spin $\ell$. In order to compute this divergent part it is therefore sufficient to study the tail of the sum in \eqref{eq:Hfun.def}. As explained in the following section such computations can be done explicitly. For example, at $\tau_0=2$ it renders
\begin{equation}\label{eq:H2.div}
H^{(0)}_2(u,v)=c\frac{u}{v}+\mathcal{O}(v^0).
\end{equation}
One notices that the power divergence agrees with the explicit calculation in \eqref{eq.H2expl}. Moreover, the finite term $\mathcal{O}(v^0)$ will not be necessary in the following sections.

Throughout the paper we will often be interested in comparing only the enhanced divergent part of various functions. For this reason we introduce a notation
\begin{equation}
f(u,v)\doteq g(u,v) \quad \mathrm{if} \quad f(u,v)=g(u,v)+\text{regular terms in the limit $ v\to0$}.
\end{equation}
Here, by the ``regular terms'' we mean contributions which can come from a finite number of conformal blocks. In particular, they can contain a single power of $\log v$ but no higher powers of the logarithm nor inverse powers of $v$. 

\subsection{Computing H-functions}
We now describe how to construct the power divergent part of the H-functions that we will need in the subsequent calculations. First, we describe how to use the kernel method, motivated by \cite{Alday:2007mf} and systematically developed in \cite{Komargodski:2012ek,Fitzpatrick:2012yx}. This method, however, becomes inefficient very fast. For this reason we explain how to use an alternative method based on the recursion relation \eqref{eq:Hrec}. We start by focusing on the case of operators with twist $\tau_0=2$, and later on describe how H-functions for higher twists arise naturally from the twist-two case.

\subsubsection{Factorisation}
We are only interested in the terms with a power divergence as $v\to 0$. In the following, it will be more convenient to use the coordinates $(\xx_1,\xx_2)$ instead of the cross-ratios $(u,v)$. In these coordinates we are interested in the limit $\xx_2\to1$. 
Using the definition \eqref{eq:Hfun.def} and the explicit form of conformal blocks, any power divergent contributions to twist conformal blocks must arise from an infinite sum over spins. Moreover, they can only come from the second part of the conformal block \eqref{eq:ConformalBlock}. Then the part of the twist conformal blocks with a power divergence as $\xx_2\to 1$ can be written as
\begin{equation}\label{eq:factorised.CBlocks}
\frac{\xx_1\,\xx_2}{\xx_2-\xx_1}k_{\frac{\tau_0}{2}-1}(\xx_1)\sum_{\ell=0}^\infty \langle a^{(0)}_{\tau_0,\ell}\rangle k_{\frac{\tau_0}{2}+\ell}(\xx_2).
\end{equation} 
Similar reasoning can be applied to all H-functions defined in \eqref{eq:HfunctionsDef}. For this reason the power divergent part of the H-functions takes a factorised form
\begin{equation}\label{eq:factorised.form}
H_{\tau_0}^{(m,\log^n)}(\xx_1,\xx_2)\doteq\frac{\xx_1}{\xx_2-\xx_1}k_{\frac {\tau_0}2-1}(\xx_1) \overline H_{\tau_0}^{(m,\log^n)}(\xx_2),
\end{equation}
where we have defined the functions 
\begin{equation}\label{eq:def.Hbar}
\overline H^{(m,\log^n)}_{\tau_0}(\xx_2)=\xx_2
\sum_{\ell=0}^\infty \langle a^{(0)}_{\tau_0,\ell}\rangle \frac{\log^nJ_{\tau_0}}{J_{\tau_0}^{2m}}\, k_{\frac{\tau_0}{2}+\ell}(\xx_2).
\end{equation}

We notice now that the action of the quadratic Casimir \eqref{eq:fullCasimir} simplifies significantly when applied only to the divergent part of the H-functions
\begin{equation}\label{eq:casimiraction.red}
\mathcal C H_{\tau_0}^{(m,\log^n)}(\xx_1,\xx_2)\doteq\frac{\xx_1}{\xx_2-\xx_1}k_{\frac{\tau_0}{2}-1}(\xx_1)\, \overline{\mathcal D} \,\overline H_{\tau_0}^{(m,\log^n)}(\xx_2),
\end{equation}
where
\begin{equation}\label{eq:Casimir.reduced}
\overline{\mathcal D}= (2-\xx_2)(1-\xx_2\partial_2)+\xx_2^2(1-\xx_2)\partial_2^2.
\end{equation}
Additionally, due to \eqref{eq:casimiraction.red}, the recursion \eqref{eq:Hrec} implies a similar recursion relation for $\overline H_{\tau_0}^{(m,\log^n)}(\xx_2)$, taking the form
\begin{equation}\label{eq:rec.simple}
\overline{H}_{\tau_0}^{(m,\log^n)}(\xx_2)=\overline{\mathcal{D}}\, \overline{H}_{\tau_0}^{(m+1,\log^n)}(\xx_2)\,.
\end{equation}

It is important to notice that the operator $\overline{\mathcal{D}}$ maps regular terms to regular terms and therefore does not introduce any enhanced divergence while acting on finite sums of conformal blocks. More generally, for polynomial functions $p(\xx_2)$ it acts as
\begin{equation}
\overline{\mathcal{D}} (p(\xx_2)\log(1-\xx_2)^n)=\frac{n(n-1)\,\xx_2\, p(\xx_2)\log(1-\xx_2)^{n-2}}{1-\xx_2}+\mathcal{O}((1-\xx_2)^0).
\end{equation}
It is clear that for $n=0,1$ no enhanced divergence is produced when acting with $\overline{\mathcal{D}}$. On the other hand, expressions with higher powers of the logarithm, namely $n>1$, will always produce terms with negative powers of $1-\xx_2$ after we act on them with $\overline{\mathcal{D}}$ a finite number of times. This property explains why we refer to such terms as enhanced divergent.

\subsubsection{Derivation of H-functions: Kernel Method}\label{sec:kernel.method}
Let us now focus on finding the power divergent part of the functions $\overline H_{\tau_0}^{(m,\log^n)}(\xx_2)$. In principle, this is possible for any $m$ and $n$. However, in order to solve the one-loop problem we will see that it is sufficient to focus on $\overline H_{\tau_0}^{(m,\log^n)}(\xx_2)$ for $n=0,1$ and $m\leq 0$. Since we want to compute just the power divergent part of these functions we only need to consider the tail of the sum over spins in \eqref{eq:def.Hbar}. In this limit the sum is well-approximated by an integral which can be explicitly computed using the method described in \cite{Komargodski:2012ek,Fitzpatrick:2012yx}, see also the appendix A of \cite{Alday:2015ewa}. This method allows to capture all power divergences, namely all terms of the form $\sim \frac{1}{(1-\xx_2)^k}$ for $k>0$. 

Let us start by considering the twist conformal block $\overline{H}_{2}^{(0)}(\xx_2)$ and compute
\begin{equation}
 \xx_2\sum_\ell \langle a_{2,\ell}^{(0)}\rangle\,k_{\ell+1}(\xx_2)=\sum_\ell 2c\frac{\Gamma(\ell+1)^2}{\Gamma(2\ell+1)}\, \xx_2^{\ell+2}\,{_2F_1}(\ell+1,\ell+1,2\ell+2;\xx_2).
\end{equation}
The divergent contributions come from large spins of order $\ell\sim\frac{1}{\sqrt{\varepsilon}}$, where we have introduced the notation $\varepsilon=1-\xx_2$ in order to simplify the following formulae. Therefore, we can define $\ell=\frac{p}{\sqrt{\varepsilon}}$ and convert the sum over $\ell$ into the integral $\frac12\int \frac{dp}{\sqrt\varepsilon}$. We also replace the hypergeometric function by its integral representation
\begin{equation}
{_2F_1}(a,b;c;\xx)=\frac{\Gamma(c)}{\Gamma(b)\Gamma(c-b)}\int\limits_0^1d t\frac{t^{b-1}(1-t)^{c-b-1}}{(1-\xx\, t)^a} .
\end{equation}
Consecutively, we perform the change of variables
\begin{equation}
\frac{p}{\sqrt{\varepsilon}}\left(\frac{p}{\sqrt{\varepsilon}}+1\right)=\frac{j^2}{\varepsilon},\qquad t=1-w\sqrt{\varepsilon}\,.
\end{equation}
 The integration limits of the $w$ integral can safely be extended to $[0,\infty)$ since this does not add any power divergent term. 
Implementing these changes of variables gives the result
\begin{equation}
 \xx_2\sum_\ell \langle a_{2,\ell}^{(0)}\rangle\,k_{\ell+1}(\xx_2)\rightarrow(1-\varepsilon)\,c \int_0^\infty dj\, \mathcal K_2(j,\varepsilon),
\end{equation}
where we have defined the integral kernel
\begin{equation}
\mathcal K_2(j,\varepsilon)=\int_0^\infty dw\,\frac{-2j}{w\,\varepsilon(w\sqrt\varepsilon-1)}\left(\frac{w(1-\varepsilon)(1-w\sqrt\varepsilon)}{w+\sqrt\varepsilon-w\varepsilon)}\right)^{\frac12\left(1+\sqrt{1+\frac{4j^2}\varepsilon}\right)}.
\end{equation}
Expanding $\mathcal K_2(j,\varepsilon)$ in powers of $\varepsilon$ we get
\begin{equation}\label{eq:expanding.kernel}
\mathcal K_2(j,\varepsilon)=4j K_0(2j)\frac1\varepsilon-\frac43\left(j K_0(2j)+(1+2j^2)K_1(2j)\right)+\ldots,
\end{equation}
where $K_n(x)$ are the modified Bessel functions of the second kind.

In particular, this method allows us to find
\begin{equation}\label{eq:kernel.H2}
\overline H^{(0)}_2(\xx_2)\doteq\frac{1}{1-\xx_2}c\int_0^\infty dj\, 4jK_0(2j)= \frac1{1-\xx_2}c+\mathcal{O}((1-\xx_2)^0),
\end{equation}
which is exactly the previously mentioned result \eqref{eq:H2.div}. Importantly, it agrees up to regular terms with the direct calculation \eqref{eq.H2expl}. Let us emphasise that for the twist conformal block $\overline{H}^{(0)}_2(\xx_2)$ there are no additional enhanced divergences beyond the power divergence, namely there are no terms with $\log^n(1-\xx_2)$ for $n>1$. This statement will become crucial when we use the recursion relation method in the following section.

More generally, using this method we can find all negative powers of $\varepsilon=1-\xx_2$ of the H-functions with $m\leq 0$ by modifying the integrand with suitable insertions
\begin{equation}\label{eq:kernel.Hbar}
\overline{H}_{2}^{(m,\log^n)}(\xx_2)\doteq
(1-\varepsilon)\,c
\int_0^\infty dj\, \mathcal K_2(j,\varepsilon)\left(\frac\varepsilon{j^2}\right)^m\log^n\left(\frac{j}{\sqrt\varepsilon}\right).
\end{equation}
For example for $m=0$, $n=1$ we find after an explicit calculation
\begin{align}\nonumber
\overline H^{(0,\log)}_2(\xx_2)&\doteq\frac{1}{1-\xx_2}c\int_0^\infty dj\, 4jK_0(2j)\left(\log j-\frac12\log(1-\xx_2)\right)\\
&\doteq -\frac{\gamma_E}{1-\xx_2}c-\frac{\log(1-\xx_2)}{2(1-\xx_2)}c+\mathcal{O}((1-\xx_2)^0),
\end{align}
where $\gamma_E$ is Euler's constant.

By studying the $\varepsilon$-dependence in \eqref{eq:kernel.Hbar} we also immediately find a general schematic form of the power divergent part of $\overline H_2^{(m,\log^n)}(\xx_2)$ for $m\leq 0$,
\begin{equation}\label{eq:form.of.Hfunctions}
\overline{H}_{2}^{(m,\log^n)}(\xx_2)\doteq \sum_{i=0}^{-m}\sum_{j=0}^{n}k_{i,j}^{(m,\log^n)}\frac{\log^{j}(1-\xx_2)}{(1-\xx_2)^{-m-i+1}}c,
\end{equation}
where all coefficient $k_{i,j}^{(m,\log^n)}$ in principle can be  calculated from \eqref{eq:kernel.Hbar}. This quickly becomes very tedious and for this reason we present a different approach in the following section.

\subsubsection{Derivation of H-functions: Recursion Relation Method}
We will now move to a more efficient approach, where we derive the H-functions $\overline H_2^{(m,\log^n)}(\xx_2)$ using the recursion relation \eqref{eq:rec.simple}. 
From \eqref{eq.H2expl} the complete enhanced divergent part of twist conformal block for $\tau_0=2$ is $\overline H_2^{(0)}(\xx_2)\doteq \frac{c}{1-\xx_2}$. The recursion relation \eqref{eq:rec.simple} immediately allows us to find all divergent parts for all H-functions $\overline H_{2}^{(m)}(\xx_2)$ with $m<0$ by simply using 
\begin{equation}\label{eq:recminusm}
\overline{H}^{(m)}_{2}(\xx_2)=\overline{\mathcal{D}}^{-m} \overline{H}_{2}^{(0)}(\xx_2)\,,\qquad \mathrm{for}\quad m<0\,.
\end{equation} 
Also for positive $m$ we could in principle find the enhanced divergent part of the H-functions by solving differential equations \eqref{eq:rec.simple}. This becomes tedious very quickly and moreover we would need to introduce two constants of integration every time we increase $m$. However, as we already pointed out, we will not need H-functions with positive $m$ at all. Left to construct are therefore the H-functions with logarithmic insertions. As described in the appendix A.4 of \cite{Alday:2016jfr}, these are given by differentiating the $\overline H^{(m)}_2(\xx_2)$ with respect to the parameter $m$:
\begin{equation}\label{eq:Hlog}
\overline H_2^{(m,\log^n)}(\xx_2)=-\frac12\frac\partial{\partial m} \overline H_2^{(m,\log^{n-1})}(\xx_2).
\end{equation}
We will only need to consider the case $n=1$, although the computation for $n>1$ is analogous. In order to find $\overline H^{(0,\log)}_2(\xx_2)$, we need to analytically continue $\overline H^{(m)}_2(\xx_2)$ with respect to the parameter $m$ and then take the derivative. The most general form of the enhanced divergent parts of $\overline H^{(m)}_2(\xx_2)$ for $m\leq 0$ is given by \eqref{eq:form.of.Hfunctions},
\begin{equation}\label{eq:Hm.exp}
\overline{H}_2^{(m)}(\xx_2)\doteq\sum_{i=0}^{-m} \frac{k^{(m)}_i}{(1-\xx_2)^{-m-i+1}}c,
\end{equation}
where all coefficients $k_i^{(m)}$ can be found explicitly from \eqref{eq:recminusm}. In particular, it allows us to derive a recursion relation for the coefficients $k_i^{(m)}$. For example for $k^{(m)}_{0}$ we get
\begin{equation}
k_0^{(m)}=m^2\,k_0^{(m+1)},
\end{equation}
which together with the initial condition $k_0^{(0)}=1$ coming from $\overline H_2^{(0)}(\xx_2)\doteq c\, (1-\xx_2)^{-1}$ allows us to find the general form
\begin{equation}
k_{0}^{(m)}=\Gamma(-m+1)^2 \,,\qquad \mathrm{for}\quad m\leq 0.
\end{equation}
Proceeding to subleading terms, and using as boundary conditions the explicit values of $k_i^{(-i)}$ for $i>0$ that can be calculated directly from \eqref{eq:recminusm}, one can find all expansion terms in \eqref{eq:Hm.exp}. We present few first terms below
\begin{align}\label{eq:Hm.exp.expl}
\overline H_2^{(m)}(\xx_2)&\doteq\frac{\Gamma(-m+1)^2c}{(1-\xx_2)^{-m+1}}+\frac{m(2m^2-6m+1)}{3}\frac{\Gamma(-m)^2c}{(1-\xx_2)^{-m}}+\nonumber\\&\quad+\frac{(m-1)m(m+1)(20m^3-54 m^2-35m+36)}{90}\frac{\Gamma(-m-1)^2c}{(1-\xx_2)^{-m-1}}+\ldots.
\end{align}
For all $m\leq 0$ this expansion is valid up to the order $(1-\xx_2)^{-1}$. Now, all expressions in \eqref{eq:Hm.exp.expl} are meromorphic functions and can be analytically continued to any value of $m$. Taking the derivative with respect to $m$, as in \eqref{eq:Hlog}, we obtain the divergent part of $\overline H_2^{(m,\log)}(\xx_2)$
\begin{align}\label{eq:Hlog.expl}
\overline H_2^{(m,\log)}(\xx_2)\doteq&-\frac{1}{2}\frac{\Gamma(-m+1)^2c}{(1-\xx_2)^{-m+1}}\left(\log(1-\xx_2)-2S_1(-m)+2\gamma_E\right)+\ldots\,,
\end{align}
where $S_k(N)=\sum_{i=1}^N \frac{1}{i^k}$ are harmonic sums. Again, for given $m\leq0$, this expansion is valid up to the order $(1-\xx_2)^{-1}$.

There exists a very compact way to encode all negative powers of $1-\xx_2$ in the functions $\overline{H}^{(m,\log)}_2(\xx_2)$ for $m\leq0$ by constructing the complete enhanced divergent part of $\overline H^{(0,\log)}_2(\xx_2)$. In order to do that we start with a general ansatz
\begin{equation}
\overline H_2^{(0,\log)}(\xx_2)=\frac{e_{\log}}{1-\xx_2}c\log(1-\xx_2)+\frac{e_{-1}}{1-\xx_2}c+\sum_{i=0}^\infty e_i(1-\xx_2)^ic\log^2(1-\xx_2).
\end{equation}
We can fix the coefficients $e_{i}$ and $e_{\log}$ by using the relation
\begin{equation}\label{eq:reclog}
\overline{H}^{(m,\log)}_{\tau_0}(\xx_2)=\overline{\mathcal{D}}^{-m} \overline{H}_{\tau_{0}}^{(0,\log)}(\xx_2)\,,\qquad \mathrm{for}\quad m<0,
\end{equation}
and comparing it with the previously obtained expansion \eqref{eq:Hlog.expl}. 
This allows us to find
\begin{equation}\label{eq:H0log}
\overline H_2^{(0,\log)}(\xx_2)=-\frac12\frac{\log(1-\xx_2)}{1-\xx_2}c-\frac{\gamma_E}{1-\xx_2}c+\left(-\frac1{12}+\frac{1-\xx_2}{10}-\frac{5(1-\xx_2)^2}{504}+\ldots\right)c\log^2(1-\xx_2).
\end{equation}
With this method arbitrarily many terms multiplying $\log^2(1-\xx_2)$ can be computed if we use~\eqref{eq:reclog} for a sufficiently large $-m$. We refer the reader to the Appendix~\ref{app:H0log} where we have collected more orders of this expansion. Now, using the explicit form of $\overline{H}_2^{(0,\log)}(\xx_2)$ in \eqref{eq:H0log} we can easily find all negative powers of $\overline{H}_2^{(m,\log)}(\xx_2)$ for $m\leq 0$ by applying the formula \eqref{eq:reclog}. A similar analysis can be done also for $\overline H^{(m,\log^n)}_2(\xx_2)$ for $n>1$, however we will not need these functions in solving the one-loop problem.

 \subsubsection{Higher Twist H-functions}

We end this section by describing how to compute the H-functions $\overline{H}_{\tau_0}^{(m,\log^n)}(\xx_2)$ for $\tau_0>2$. First of all, notice that the tree-level structure constants for higher twists \eqref{eq:structureconstants} can be nicely written using the tree-level structure constants for twist-two operators
\begin{equation}
\langle a^{(0)}_{\tau_0,\ell}\rangle=\frac{\Gamma(\frac{\tau_0}{2}-1)^2}{\Gamma(\tau_0-3)}\frac{1}{c}\left(c\,(-1)^{\frac{\tau_0}{2}}-(\tfrac{\tau_0}{2}-2)(\tfrac{\tau_0}{2}-1)+J_{\tau_0}^2\right)\langle a^{(0)}_{2,\ell+\frac{\tau_0}{2}-1}\rangle,
\end{equation}
where again $J_{\tau_0}^2=\left(\frac{\tau_0}{2}+\ell\right)\left(\frac{\tau_0}{2}+\ell-1\right)$.
When we plug this into the definition of twist conformal blocks for higher twist and perform a change of variables $\ell'=\ell+\frac{\tau_0}{2}-1$ we get
\begin{equation}
\overline{H}^{(0)}_{\tau_0}(\xx_2)=\xx_2\frac{\Gamma(\frac{\tau_0}{2}-1)^2}{\Gamma(\tau_0-3)}\sum_{\ell'=\frac{\tau_0}{2}-1}^\infty \frac{1}{c}\left(c\,(-1)^{\frac{\tau_0}{2}}-(\tfrac{\tau_0}{2}-2)(\tfrac{\tau_0}{2}-1)+(J'_{2})^2\right)\langle a^{(0)}_{2,\ell'}\rangle k_{\ell'+1}(\xx_2).
\end{equation}
where $(J'_2)^2=\ell'(\ell'+1)$.
 In the limit $\xx_2\to1$ the sum over $\ell'$ can be replaced by a sum from zero to infinity since the difference is a regular term. This leads to
\begin{equation}
\overline H^{(0)}_{\tau_0}(\xx_2)\doteq\frac{\Gamma(\frac{\tau_0}{2}-1)^2}{\Gamma(\tau_0-3)}\frac{1}{c}\left(\left(c\,(-1)^{\frac{\tau_0}{2}}-(\tfrac{\tau_0}{2}-2)(\tfrac{\tau_0}{2}-1)\right)\overline H^{(0)}_2(\xx_2)+\overline H_2^{(-1)}(\xx_2)\right).
\end{equation}
This allows us to rewrite the twist conformal blocks for higher twists in terms of functions we have already constructed. Similar analysis can be performed for all H-functions leading to the explicit form for higher-twists 
\begin{equation}\label{eq:computing.higher.twist.H}
\overline H^{(m,\log^n)}_{\tau_0}(\xx_2)\doteq\frac{\Gamma(\frac{\tau_0}{2}-1)^2}{\Gamma(\tau_0-3)}\frac{1}{c}\!\left(\!\left(c\,(-1)^{\frac{\tau_0}{2}}-(\tfrac{\tau_0}{2}-2)(\tfrac{\tau_0}{2}-1)\right)\!\overline H^{(m,\log^n)}_2(\xx_2)+\overline H^{(m-1,\log^n)}_2(\xx_2)\right)\!.
\end{equation}
To summarise, all H-functions relevant for the one-loop problem can be constructed using just two functions: $\overline H^{(0)}_2(\xx_2)$ and $\overline H^{(0,\log)}_2(\xx_2)$ whose explicit form can be found in \eqref{eq:H2.div} and \eqref{eq:H0log}, respectively.

\subsection{Decomposing One-Loop Correlator into H-functions}
\label{sec:one-loop.correlator.into.H}
Knowing the explicit form of the H-functions, we focus now on the one-loop four-point correlation function $\mathcal{G}^{(1)}(\xx_1,\xx_2)$ and expand its power divergent part in terms of the H-functions. By doing this we focus only on  contributions to anomalous dimensions and structure constants unbounded in spin $\ell$. Later on we will also include terms which are truncated in spin. The latter do not interfere with our analysis of the power divergent part of the correlator. 

For each operator present in the intermediate channel we expand their conformal dimension and structure constants as follows
\begin{align}
\tau_i&=\tau_0+g\, \gamma^{(1)}_{\tau_0,\ell,i}+\mathcal{O}(g^2),\\ a_{\tau_i,\ell,i}&=a^{(0)}_{\tau_0,\ell,i}+g\, a_{\tau_0,\ell,i}^{(1)}+\mathcal{O}(g^2).
\end{align}
Then the four-point correlation function $\mathcal{G}(\xx_1,\xx_2)$, up to the order $g$, can be written as
\begin{align}\label{eq:G0G1exp}
&\mathcal{G}^{(0)}(\xx_1,\xx_2)+g\, \mathcal{G}^{(1)}(\xx_1,\xx_2)\nonumber\\
&=\sum\limits_{\tau_0,\ell,i}\left(a^{(0)}_{\tau_0,\ell,i}+g\, a^{(1)}_{\tau_0,\ell,i}\right)\left(G_{\tau_0,\ell}(\xx_1,\xx_2)+g\,\gamma^{(1)}_{\tau_0,\ell,i}\left(\frac\partial{\partial\tau} G_{\tau,\ell}(\xx_1,\xx_2)\right)\big|_{\tau\to\tau_0}\right)
\\
&=\sum\limits_{\tau_0,\ell}\langle a_{\tau_0,\ell}^{(0)}\rangle G_{\tau_0,l}(\xx_1,\xx_2)+g\,\sum\limits_{\tau_0,\ell}\left(\langle a_{\tau_0,\ell}^{(1)}\rangle G_{\tau_0,\ell}(\xx_1,\xx_2)+\langle a_{\tau_0,\ell}^{(0)}\gamma^{(1)}_{\tau_0,\ell}\rangle\left(\frac\partial{\partial\tau} G_{\tau,\ell}(\xx_1,\xx_2)\right)\big|_{\tau\to\tau_0}\right)\nonumber,
\end{align}
where we have again defined the averages $\langle f_{\tau_0,\ell}\rangle=\sum_i f_{\tau_0,\ell,i}$. 

In the last line of \eqref{eq:G0G1exp} the derivative with respect to twist $\tau$ is understood as a partial derivative of a function of two variables: $\tau$ and $\ell$. It turns out that our further analysis simplifies significantly if we instead use the variables $(\tilde\tau,\tilde\ell)$ defined as 
\begin{equation}
 \left(\tilde\tau,\,\tilde\ell\,\right)=\left(\tau,\ell+\frac{\tau}{2}\right).
\end{equation} 
Then the partial derivatives in the new variables can be related to the partial derivatives with respect to the twist and spin as
\begin{equation}
\frac\partial{\partial\tau}=\frac\partial{\partial\tilde\tau}+\frac{1}{2}\frac\partial{\partial\tilde\ell}\,,\qquad\frac\partial{\partial\ell}=\frac\partial{\partial\tilde\ell}\,.
\end{equation}
In particular, it implies that $\partial_{\tilde\tau}k_{\frac\tau2+\ell}(\xx)=0$. 
We can now rewrite the derivative in the last line of \eqref{eq:G0G1exp} as
\begin{align}
\sum\limits_{\tau_0,\ell}\left(\langle a_{\tau_0,\ell}^{(0)}\gamma^{(1)}_{\tau_0,\ell}\rangle\left(\frac\partial{\partial\tilde\tau} G_{\tau,\ell}(\xx_1,\xx_2)\right)\big|_{\tau\to\tau_0}+\frac{1}{2}\langle a_{\tau_0,\ell}^{(0)}\gamma^{(1)}_{\tau_0,\ell}\rangle\left(\frac\partial{\partial\tilde\ell} G_{\tau_0,\ell}(\xx_1,\xx_2)\right)\right)\nonumber\\
\doteq \sum\limits_{\tau_0,\ell}\left(\langle a_{\tau_0,\ell}^{(0)}\gamma^{(1)}_{\tau_0,\ell}\rangle\left(\frac\partial{\partial\tilde\tau} G_{\tau,\ell}(\xx_1,\xx_2)\right)\big|_{\tau\to\tau_0}-\frac{1}{2}\frac{\partial}{\partial \tilde\ell}\left(\langle a_{\tau_0,\ell}^{(0)}\gamma^{(1)}_{\tau_0,\ell}\rangle\right)G_{\tau_0,\ell}(\xx_1,\xx_2)\right),
\end{align}
where in the second line we dropped a total derivative with respect to $\tilde\ell$, which is a regular term. Finally, we can rewrite the divergent part of $\mathcal{G}^{(1)}(\xx_1,\xx_2)$ as
\begin{equation}\label{eq:G1.almostH}
\mathcal{G}^{(1)}(\xx_1,\xx_2)\doteq\frac{ \xx_1\xx_2}{\xx_2-\xx_1}\sum\limits_{\tau_0,\ell}\langle a^{(0)}_{\tau_0,\ell}\rangle\left(\overline{\langle \hat\alpha_{\tau_0,\ell}\rangle} k_{\frac{\tau_0}{2}-1}(\xx_1)+\overline{\langle\gamma_{\tau_0,\ell}\rangle}\left(\frac\partial{\partial\tau} k_{\frac{\tau}{2}-1}(\xx_1)\right)\big|_{\tau\to\tau_0} \right)k_{\frac{\tau_0}{2}+\ell}(\xx_2),
\end{equation}
where we used the factorisation~\eqref{eq:factorised.CBlocks} of the divergent parts of the conformal blocks and introduced
\begin{align}\label{eq:gammatau}
\overline{\langle  \gamma_{{\tau_0},\ell}\rangle} &\equiv \frac{\langle  a^{(0)}_{{\tau_0},\ell}\gamma^{(1)}_{{\tau_0},\ell}\rangle}{\langle  a^{(0)}_{{\tau_0},\ell}\rangle},\\
\overline{\langle \hat \alpha_{{\tau_0},\ell}\rangle}
\langle a_{{\tau_0},\ell}^{(0)}\rangle
&\equiv
 \langle a_{{\tau_0},\ell}^{(1)}\rangle-\frac{1}{2}\frac\partial{\partial\ell}\left(\langle a^{(0)}_{{\tau_0},\ell}\gamma^{(1)}_{{\tau_0},\ell}\rangle\right)=\langle a_{{\tau_0},\ell}^{(1)}\rangle
 -
 \frac{1}{2}\langle a^{(0)}_{{\tau_0},\ell}\rangle\frac\partial{\partial\ell}\overline{\langle \gamma_{{\tau_0},\ell}\rangle}
 -
 \frac{1}{2}\frac\partial{\partial\ell}\langle a^{(0)}_{{\tau_0},\ell}\rangle\overline{\langle \gamma_{{\tau_0},\ell}\rangle}.\label{eq:modified.structure.constant}
\end{align}
One can recognise the last formula in \eqref{eq:modified.structure.constant} as the one-loop perturbative expansion of $\hat a_{\tau_0,\ell}$ introduced in \cite{Alday:2016jfr}.

In weakly coupled CFTs at one loop, both the anomalous dimensions $\overline{\langle  \gamma_{{\tau_0},\ell}\rangle}$ and the modified structure constants $\overline{\langle \hat \alpha_{{\tau_0},\ell}\rangle}$ depend on spin as a single logarithm $\log \ell$ at large $\ell$. Therefore, in order to use the H-functions to constrain the unbounded parts of the CFT data we expand the modified structure constants $\overline{\langle\hat\alpha_{\tau_0,\ell}\rangle}$ and anomalous dimensions $\overline{\langle\gamma_{\tau_0,\ell}\rangle}$ in the following way \cite{Alday:2015eya}:
\begin{align}\label{eq:OPEexpansion}
\overline{\langle \hat \alpha_{\tau_0,\ell}\rangle} &=\sum_{m=0}^\infty \frac{A_{\tau_0,(m,\log)}}{J^{2m}_{\tau_0}}\log J_{\tau_0}+\sum_{m=0}^\infty \frac{A_{\tau_0,(m)}}{J_{\tau_0}^{2m}},\quad\\\label{eq:gammaexpansion}
\overline{\langle  \gamma_{\tau_0,\ell}\rangle} &=\sum_{m=0}^\infty \frac{B_{\tau_0,(m,\log)}}{J^{2m}_{\tau_0}}\log J_{\tau_0}+\sum_{m=0}^\infty \frac{B_{\tau_0,(m)}}{J_{\tau_0}^{2m}}.
\end{align}

Inserting the expansions \eqref{eq:OPEexpansion} and \eqref{eq:gammaexpansion} into \eqref{eq:G1.almostH} we can finally rewrite the divergent part of the one-loop correlator in terms of H-functions
\begin{align}\label{eq:oneloop.in.Hfunctions}
\mathcal G^{(1)}(\xx_1,\xx_2)\doteq \sum_{\tau_0}\frac{ \xx_1}{\xx_2-\xx_1}\sum_\rho\left( A_{\tau_0,\rho}\,k_{\frac{\tau_0}{2}-1}(\xx_1)+B_{\tau_0,\rho}\,\left(\frac\partial{\partial\tau} k_{\frac{\tau}{2}-1}(\xx_1)\right)\big|_{\tau\to\tau_0}\right) \overline H_{\tau_0}^{\,\rho}(\xx_2),
\end{align}
where $\rho=(m,\log)$ or $\rho=(m)$, $m=0,1,2,\ldots$ and we have used the definition of H-functions \eqref{eq:def.Hbar}. This is the most important formula of this section and in the following we will use it to completely fix the form of $\mathcal{G}^{(1)}(\xx_1,\xx_2)$. 

\subsection{Using H-functions: Toy Example}
\label{sec:method}

We present a simple example of how to use H-functions to extract the asymptotic spin dependence of CFT data given a particular function with power divergences. In order to simplify our discussion we focus here only on the $\xx_2$ dependence. In analogy with the actual computations in the next section, we will assume that the sum of H-functions produces a divergent expression containing a constant term and a term proportional to $\log(1-\xx_2)$:
\begin{equation}\label{eq:toy.equation}
\sum_{m=0}^\infty\sum_{n=0}^1 C_{(m,\log^n)}\overline H_2^{(m,\log^n)}(\xx_2)\doteq\frac{\lambda_1\log(1-\xx_2)+\lambda_0}{1-\xx_2}c.
\end{equation}
We will work iteratively and fix coefficients $C_{(m,\log^n)}$ by repeatedly applying the Casimir operator \eqref{eq:Casimir.reduced} on both sides of \eqref{eq:toy.equation} and keeping only power divergent terms. As a first step let us analyse the power divergent terms of \eqref{eq:toy.equation} itself. In this case only two terms in the sum on the left hand side are power divergent as $\xx_2\to1$, namely $\overline H^{(0)}_2(\xx_2)$ and $\overline H^{(0,\log)}_2(\xx_2)$. Therefore we get
\begin{equation}
C_{(0)}\frac{1}{1-\xx_2}c+C_{(0,\log)}\left( -\frac{\gamma_E}{1-\xx_2}-\frac{\log(1-\xx_2)}{2(1-\xx_2)}\right)c =
\frac{\lambda_1\log(1-\xx_2)+\lambda_0}{1-\xx_2}c,
\end{equation}
where we used the explicit form of $\overline H^{(0)}_2(\xx_2)$ and  $\overline H^{(0,\log)}_2(\xx_2)$. Solving this equation we get
\begin{equation}\label{eq:toy.zero.order.sols}
C_{(0)}=\lambda_0-2\lambda_1\gamma_E,\qquad C_{(0,\log)}=-2\lambda_1.
\end{equation}
To compute higher coefficients we act with the Casimir $\overline{\mathcal D}$ on both sides of \eqref{eq:toy.equation} and again compare power divergent terms. On the left hand side, using the recurrence \eqref{eq:rec.simple}, the Casimir brings the previously undetermined coefficients $C_{(1)}$ and $C_{(1,\log)}$ into the problem. This renders
\begin{equation}
\sum_{m=0}^1 \sum_{n=0}^1 C_{(m,\log^n)}\overline H^{(m-1,\log^n)}_2(\xx_2) \doteq\overline{\mathcal D}\left(\frac{\lambda_1\log(1-\xx_2)+\lambda_0}{1-\xx_2}c \right).
\end{equation}
Using the explicit form of the H-functions
\begin{align}
\overline H^{(-1)}_2(\xx_2)&=\overline{\mathcal{D}}\,\overline{H}^{(0)}_2(\xx_2)\doteq\frac{1}{(1-\xx_2)^2}c-
\frac{3}{1-\xx_2}c,\\\!\!
\overline H^{(-1,\log)}_2(\xx_2)&=\overline{\mathcal{D}}\,\overline{H}^{(0,\log)}_2(\xx_2)\doteq
\frac{2-2\gamma_E-\log(1-\xx_2)}{2(1-\xx_2)^2}c
+
\frac{18\gamma_E-19+9\log(1-\xx_2)}{6(1-\xx_2)}c,\!
\end{align}
and plugging in the solutions \eqref{eq:toy.zero.order.sols}, the term proportional to $(1-\xx_2)^{-2}$ vanishes, and the term proportional to $(1-\xx_2)$ provides
\begin{equation}
C_{(1)}=-\frac{\lambda_1}{3},\qquad C_{(1,\log)}=0.
\end{equation}
We can continue in this fashion, and determine the coefficients $C_{(m)}$ and $C_{(m,\log)}$ after acting $m$ times with the Casimir $\overline{\mathcal D}$. The results for $m=1,2,\ldots$ are
\begin{equation}
C_{(m)}=-2\lambda_1\left\{\frac{1}{6},\,-\frac{1}{30},\,\frac{4}{315},\,-\frac{1}{105},\,\ldots\right\},\qquad C_{(m,\log)}=0.
\end{equation}
We can identify the $C_{(m)}$ together with $C_{(0,\log)}$ as coefficients in the large $\ell$ expansion of the harmonic sum  $S_1(\ell)$ expanded in inverse powers of $J^2=\ell(\ell+1)$. They therefore describe a function
\begin{equation}
\sum_{m,n} {C_{(m,\log^n)}}\frac{\log^n J}{J^{2m}}=\lambda_0-2\lambda_1S_1(\ell).
\end{equation}
This computation proves the following relation, which can also be shown by explicit computation,
\begin{equation}\label{eq:toy.example.starting.point}
\sum_\ell \langle a_{2,\ell}^{(0)}\rangle\, \xx_2\,k_{\ell+1}(\xx_2)( \lambda_0-2\lambda_1S_1(\ell)) \doteq \frac{\lambda_1\log(1-\xx_2)+\lambda_0}{1-\xx_2}c.
\end{equation}

In the following we will apply this method to more complicated functions, but the general idea will stay exactly the same.

\section{Four-Point Correlator from H-functions}
\label{sec:FindingNemo}

In this section we use the H-functions to construct the one-loop correction to the four-point function of four identical scalar operators. Again, we think of the correlator of four Konishi operators as our example, but the method applies to a large family of scalar operators. 

\subsection{The Strategy}
We remind the reader that the four-point correlation function in weakly coupled gauge theories admits an expansion in the coupling constant $g$ of the form
\begin{equation}
\mathcal{G}(u,v)=\mathcal{G}^{(0)}(u,v)+g \,\mathcal{G}^{(1)}(u,v)+\ldots.
\end{equation}
The contributions to the one-loop correlator $\mathcal{G}^{(1)}(u,v)$ come from two different sources. First of all, there are infinite towers of operators for which the CFT data can be expanded as a power series at large spin $\ell$, with possible $\log \ell$ insertions. Such towers of operators necessarily produce power divergent contributions to the correlator and we can study them using the H-functions. Secondly, there are terms in the four-point correlator which after performing the conformal partial wave decomposition render CFT data that is truncated in spin. Such terms are always regular as $v\to0$. Importantly, these two kinds of contributions mix under the crossing. 
In fact, the mixing is such that all contributions from infinite towers, at any twists, are completely determined by the twist-two operators. 
Therefore we will start our analysis from general ansatz for the twist-two operators, and then use the crossing symmetry and the H-function method to extend the ansatz to a full solution for the one-loop four-point correlator. 
In the process we will assume that there are no truncated solutions of the form found in \cite{Heemskerk:2009pn}. 

Our strategy to find the one-loop correlation function is the following:
\begin{itemize}
 \item Using the explicit form of conformal blocks \eqref{eq:ConformalBlock} and the bootstrap equation \eqref{crossing.symmetry} we find a general form of the power divergent part of $\mathcal{G}^{(1)}(u,v)$ in the limit $v\to0$. We show using crossing symmetry that this is fully described by operators at leading twist, namely $\tau_0=2$. Subsequently, we use the H-function method to constrain the form of the contributions from infinite towers of leading twist operators. Supplementing this with terms truncated in spin we arrive at the most general leading twist contribution to the correlator $\mathcal G_{\mathrm{L.T.}}(u,v)\sim uf(\log u,v)$, where $f(\log u,v)$ is expressed to all orders in $v$ in terms of a finite number of unknowns.
\item Crossing symmetry maps $u f(\log u,v)$ to the power divergent part of the complete four-point correlator. This allows us to use the H-function method to find the large spin expansion of the CFT data for all twists, which can be resummed to  closed-form functions of spin. Plugging this result back to the conformal partial wave expansion we find the complete form of the four-point correlator in terms of a finite number of unknowns.
\item   As a final step we check that such obtained function satisfy all necessary constraints. In particular, consistency with the bootstrap equation reduces the number of unknowns to just four. 
\end{itemize}

\subsection{The Ansatz}\label{sec:ansatz}
We focus first on the most general form of the power divergent terms in the limit $v\to 0$ and show that the bootstrap equation implies that all such contributions are encoded by the twist-two operators. 

Let us start by writing down an explicit form of the bootstrap equation in the perturbative expansion 
 \begin{equation}\label{eq:crossing.Konishi}
 v^{2+g \,\gamma_{\mathrm{ext}}}(\mathcal{G}^{(0)}(u,v)+g\, \mathcal{G}^{(1)}(u,v))=u^{2+g\, \gamma_{\mathrm{ext}}}(\mathcal{G}^{(0)}(v,u)+g\, \mathcal{G}^{(1)}(v,u)),
\end{equation}
where $\gamma_{\mathrm{ext}}$ is the one-loop anomalous dimension of the external operators, which we at the moment will keep unspecified.
The one-loop part of this equation can be written in the form
\begin{equation}\label{eq:crossing.Konishi.simpl}
\tilde{\mathcal G}^{(1)}(u,v)=\frac{u^2}{v^2} \tilde{\mathcal G}^{(1)}(v,u),
\end{equation}
where for convenience we defined $\tilde{\mathcal{G}}^{(1)}(u,v)=\mathcal{G}^{(1)}(u,v)+\gamma_\mathrm{ext}\log v\,\mathcal{G}^{(0)}(u,v)$.
Both functions $\mathcal{G}^{(0)}(u,v)$ and $\mathcal{G}^{(1)}(u,v)$ can be expanded in conformal blocks. Let us then look at the expansion of a single conformal block in the small $g$ limit,
\begin{equation}\label{eq:leadingconfblock}
G_{\tau,\ell}(u,v)= G_{\tau_0,\ell}(u,v)+g\,(\partial_{\tau} G_{\tau,\ell}(u,v))|_{\tau\to\tau_0}+\mathcal{O}(g^2).
\end{equation} 
From the explicit form of the conformal blocks we notice that at one loop there is a contribution proportional to $\log u$ in this expansion but no higher powers of the logarithm. We also notice that in the small $u$ limit we have $G_{\tau_0,\ell}(u,v)\sim u^{\tau_0/2}$. Thus the first non-trivial part of $\tilde{\mathcal{G}}^{(1)}(u,v)$ at small $u$ comes exclusively from the twist-two operators and is of the form
\begin{equation}\label{eq:leadingG}
\tilde{\mathcal{G}}^{(1)}(u,v)= \gamma_{\mathrm{ext}}\log v+ u\left(Q^{(1)}(v,\log v)\log u+Q^{(2)}(v,\log v)\right)+\mathcal{O}(u^2),
\end{equation}
where the first trivial term comes from the identity operator contribution to $\mathcal{G}^{(0)}(u,v)$ and $Q^{(i)}(v,\log v)$ are arbitrary functions.
The bootstrap equation \eqref{eq:crossing.Konishi.simpl} used for \eqref{eq:leadingG} now gives
 \begin{equation}\label{eq:leadingG2}
 \tilde{\mathcal{G}}^{(1)}(u,v)=\gamma_{\mathrm{ext}}\frac{u^2}{v^2}\log u +\frac{u^2}{v}\left(Q^{(1)}(u,\log u)\log v+Q^{(2)}(u,\log u)\right)+\mathcal{O}(v^0).
 \end{equation}
We notice in particular that, when crossed, \eqref{eq:leadingG} produces a power divergence for $v\to0$. It is easy to see that also the opposite statement is true: any divergent part of $\tilde{\mathcal{G}}^{(1)}(u,v)$ is mapped to the first two leading $u$ powers under crossing. Finally, by comparing the formulae \eqref{eq:leadingG} and \eqref{eq:leadingG2} we conclude that we must have $Q^{(i)}(u,\log u)\sim \frac{1}{u}+\ldots$.

Since the term proportional to $u^0$ is completely determined by the tree-level, we will focus here on the term proportional to $u$. Therefore, we start our analysis by considering the most general ansatz for twist-two operators. There are two distinguished terms: the contributions containing a power divergent part at $v\to0$, and contributions truncated in the spin. From the discussion above, we conclude that the former takes the form
\begin{equation}\label{Ginf.leading}
 \mathcal{G}_{\mathrm{inf,L.T.}}^{(1)}(u,v)\sim\frac{u}{v} \left(\alpha_{11}\log u \log v+\alpha_{10}\log u+\alpha_{01}\log v+\alpha_{00}\right)c+\ldots\,.
\end{equation}
where $\alpha_{00},\alpha_{10},\alpha_{01},\alpha_{11}$ are arbitrary constants and we introduced an explicit dependence on $c$ for later convenience. In the subsequent part of this section, we will use the H-function method to extend this to all subleading orders in $v$.
 
For the truncated contributions, let us take $L$ such that
\begin{equation}
\begin{cases}
\langle a^{(1)}_{2,\ell}\rangle=\langle a^{(1)}_{2,\ell}\rangle_{\mathrm{inf}}+\langle a^{(0)}_{2,\ell}\rangle\mu_{\ell}\,,\\
\overline{\langle \gamma_{2,\ell} \rangle}=\overline{\langle \gamma_{2,\ell} \rangle}_{\mathrm{inf}}+\nu_\ell\,,
\end{cases}
\qquad 
\ell=0,2,\ldots,L\,,
\end{equation}
and that for spins $\ell>L$ we have only contributions from infinite towers of operators.
 In this case the truncated part of the one-loop answer is given by
\begin{equation}\label{eq.Gfin.leading}
\mathcal{G}^{(1)}_{\mathrm{trunc,L.T.}}(u,v)=\sum_{\ell=0}^L\langle a^{(0)}_{2,\ell}\rangle\left( \mu_\ell \,G_{2,\ell}(u,v)+\nu_\ell \left(\partial_{\tau} G_{\tau,\ell}(u,v)\right)\big|_{\tau\to2}  \right).
\end{equation}

Let us go back to the term containing a divergence as $v\to 0$ in \eqref{Ginf.leading}. It originates purely from an infinite tower of twist-two operators and can be expanded using H-functions as in \eqref{eq:oneloop.in.Hfunctions}:
\begin{align}\nonumber
&\frac{\xx_1}{1-\xx_2}\left( \alpha_{11}\log \xx_1 \log(1-\xx_2)+\alpha_{10}\log \xx_1 +\alpha_{01}\log(1-\xx_2) +\alpha_{00}\right)c\doteq\\
&\hspace{7cm}\doteq \xx_1 \sum_\rho\left(A_{2,\rho}+\frac{1}{2}B_{2,\rho}\log \xx_1 \right) \overline H_2^{(\rho)}(\xx_2)\,,
\end{align}  
 where $A_{2,\rho}$ and $B_{2,\rho}$ are large-$J$ expansion coefficients, as in \eqref{eq:OPEexpansion} and \eqref{eq:gammaexpansion}, of the modified structure constants and anomalous dimensions, respectively, with $\rho=(m,\log^n)$ for $n=0,1$ and $m=0,1,\ldots$. Using the H-function method described in section \ref{sec:method} we find 
 \begin{align}
 A_{2,(0,\log)}&=-2\alpha_{01},&  A_{2,(0)}&=-2\alpha_{01}\gamma_E+\alpha_{00},& A_{2,(m)} &=-2\alpha_{01}\left\{  
 \frac{1}6,\, \frac{-1}{30},\, \frac{4}{315},\,\ldots
 \right\},
 \\ 
 B_{2,(0,\log)}&=-4\alpha_{11}, &  B_{2,(0)}&=-4\alpha_{11}\gamma_E+2\alpha_{10}, &B_{2,(m)}&=-4\alpha_{11}\left\{ \frac{1}6,\, \frac{-1}{30},\, \frac{4}{315},\,\ldots \right\}.
 \end{align}
From these values we can find an explicit form of the anomalous dimension and one-loop structure constants coming from an infinite tower of twist-two operators:
 \begin{align}\label{eq:twisttwogammaa}
\overline{ \langle \gamma_{2,\ell}\rangle}_{\mathrm{inf}} &=-4\alpha_{11}\,S_1(\ell)+2\alpha_{10}\,,\\\label{eq:twisttwoalphaa}
\overline{ \langle \hat{\alpha}_{2,\ell}\rangle}_{\mathrm{inf}}&=-2\alpha_{01}\,S_1(\ell)+\alpha_{00}\,.
 \end{align}

In the next step we will take the results \eqref{eq:twisttwogammaa}, \eqref{eq:twisttwoalphaa} and plug them into the conformal partial wave expansion \eqref{eq:BlockDecomposition}. We can perform a resummation of the complete leading $\xx_1$ expansion of the four-point correlator $\mathcal{G}^{(1)}_{\mathrm{inf,L.T.}}(u,v)$ and arrive at
\begin{equation}\label{eq:Ginfleading}
\mathcal{G}^{(1)}_{\mathrm{inf,L.T.}}(u,v)=\xx_1\xx_2 \left(\alpha_{11} F_{11}( \xx_1,\xx_2)+\alpha_{10} F_{10}(\xx_1,\xx_2)+\alpha_{01} F_{01}( \xx_1,\xx_2)+\alpha_{00} F_{00}(\xx_1,\xx_2)\right),
\end{equation}  
where
\begin{align}
F_{11}(\xx_1,\xx_2)&=c\,\frac{\xx_2}{1-\xx_2}\log(1-\xx_2)\log(\xx_1\xx_2)+2c\left(\frac{\xx_2}{1-\xx_2} \mathrm{Li}_2(\xx_2)-\frac{2-\xx_2}{1-\xx_2}\zeta_2 \right),
\\
F_{10}(\xx_1,\xx_2)&=c\left( \frac1{1-\xx_2}+1 \right)\log(\xx_1\xx_2)-c\log(1-\xx_2)\,,
\\
F_{01}(\xx_1,\xx_2)&=c\left( \frac1{1-\xx_2}-1 \right)\log(1-\xx_2)\,,
\\
F_{00}( \xx_1,\xx_2)&=c\left( \frac1{1-\xx_2}+1 \right).
\end{align}
Here $\zeta_2=\frac{\pi^2}{6}$ and $\mathrm{Li}_2(\xx)$ is the dilogarithm. It is easy to confirm that the power divergent part of \eqref{eq:Ginfleading} indeed equals \eqref{Ginf.leading}. We emphasize that the expansion \eqref{eq:Ginfleading} is valid only at the leading order in $\xx_1\to0$ but is exact to all orders in $\xx_2$.

We add together \eqref{eq.Gfin.leading} and \eqref{eq:Ginfleading} to get the most general form of the one-loop correlator at the leading order in $u\to0$ expansion
\begin{equation}\label{eq:G.leading.twist}
\mathcal G^{(1)}_{\mathrm{L.T.}}(u,v)=\mathcal G^{(1)}_{\mathrm{inf,L.T.}}(u,v)+\mathcal G^{(1)}_{\mathrm{trunc,L.T.}}(u,v).
\end{equation}
This answer depends on $2L+4$ unspecified coefficients and concludes the first step in our strategy.

\subsection{Higher twist operators}
In the next step we will use the complete form of the leading twist four-point function $G^{(1)}_{\mathrm{L.T.}}(u,v)$ together with the crossing equation to study implications for higher twist operators. As we already have pointed out, the term proportional to $ u$ are, apart from the trivial contribution from the identity operator, the only ones which can produce power divergent terms after the crossing. It implies that after we apply the crossing symmetry to the function \eqref{eq:G.leading.twist} we get the complete power divergence of the full one-loop answer. 

In order to make our results more transparent, let us assume at the moment that $L=0$, namely only spin $\ell=0$ contributes to the truncated ansatz \eqref{eq.Gfin.leading}. We will come back to the general case later. Let us look again at the crossing equation \eqref{eq:crossing.Konishi} at order $g$, which gives the following equation for the one-loop correlation function:
\begin{equation}\label{eq:crossing.for.HT}
\mathcal G^{(1)}(u,v)=\frac{u^2}{v^2} \mathcal G^{(1)}(v,u)+\gamma_{\mathrm{ext}}\mathcal G^{(0)}(u,v)\left(\log u-\log v\right).
\end{equation}
From our previous computations, on the right hand side we know explicitly all power divergent contributions in the limit $v\to0$. First of all, we can expand \eqref{eq:crossing.for.HT} at leading $v\to0$ and $u\to0$ to get
 \begin{align}\nonumber
 \mathcal{G}^{(1)}(\xx_1,\xx_2)\sim\frac{\xx_1 }{(1-\xx_2)} \left(\alpha_{11}\log \xx_1 \log (1-\xx_2)+(\alpha_{01}+\gamma_{\mathrm{ext}})\log \xx_1\right.\\ \left.+(\alpha_{10}-\gamma_{\mathrm{ext}})\log (1-\xx_2)+\alpha_{00}\right)c+\ldots.
 \end{align}
 Comparing it with \eqref{Ginf.leading} we find the constraint 
 \begin{equation}
 \alpha_{01}=\alpha_{10}-\gamma_{\mathrm{ext}}.
 \end{equation}
 After substituting this into \eqref{eq:crossing.for.HT} we notice that the divergent part of $\mathcal{G}^{(1)}(u,v)$ depends on the anomalous dimension of external operator $\gamma_{\mathrm{ext}}$ and the five parameters $(\alpha_{11},\alpha_{10},\alpha_{00},\mu_0,\nu_0)$. We use this function to find the unbounded CFT data for higher twist operators by solving \eqref{eq:oneloop.in.Hfunctions}. Applying the method explained in section \ref{sec:method} we can compute as many coefficients $A_{\tau_0,(m,\log^k)}$ and $B_{\tau_0,(m,\log^k)}$ as necessary. Similar to the case of twist-two operators, we plug it back to \eqref{eq:OPEexpansion}, \eqref{eq:gammaexpansion} and we are able to perform the sum to find an explicit form of the CFT data coming from infinite towers of operators as a function of spin. The result for the anomalous dimensions is 
\begin{align}\label{eq:gamma.HT}
\overline{\langle\gamma_{\tau_0,\ell}\rangle}&=\frac{c}{P_{\tau_0,\ell}}\Big(
4\alpha_{11}\eta\left[S_1(\tfrac{\tau_0}2-2)+S_1(\tfrac{\tau_0}2+\ell-1)+\tfrac12\delta_{\tau_0,4}\right]-4\eta\,\alpha_{10}+2\eta\,\gamma_{\mathrm{ext}}
\nonumber\\
&\qquad\qquad-4\mu_0-4\nu_0\left[S_1(\tfrac{\tau_0}2-2)-S_1(\tfrac{\tau_0}2+\ell-1)+1\right]
\Big)
+2\gamma_{\mathrm{ext}},\quad \mathrm{for}\,\,\tau_0>2,
\end{align}
where $\eta=(-1)^{\frac{\tau_0}2}$ and $P_{\tau_0,\ell}=c\, \eta+(\tau_0+\ell -2)(\ell+1)$ is the factor that appears in the tree level structure constants~\eqref{eq:structureconstants}.
The result for $\overline{\langle\hat\alpha_{\tau_0,\ell}\rangle}
$ is more involved and we present here only its schematic form
\begin{equation}\label{eq:structure.constants.sum}
\overline{\langle\hat\alpha_{\tau_0,\ell}\rangle}=\alpha_{11}\overline{\langle\hat\alpha_{\tau_0,\ell}\rangle}_{11}+\alpha_{10}\overline{\langle\hat\alpha_{\tau_0,\ell}\rangle}_{10}+\alpha_{00}\overline{\langle\hat\alpha_{\tau_0,\ell}\rangle}_{00}+\gamma_{\mathrm{ext}}\overline{\langle\hat\alpha_{\tau_0,\ell}\rangle}_{\mathrm{ext}}+\mu_{0}\overline{\langle\hat\alpha_{\tau_0,\ell}\rangle}_{\mu_0}+\nu_{0}\overline{\langle\hat\alpha_{\tau_0,\ell}\rangle}_{\nu_0}.
\end{equation}
The explicit expressions for $\overline{\langle\hat\alpha_{\tau_0,\ell}\rangle}_{i}$ can be found in the appendix~\ref{app:results.nonsuper}. In order to get the one-loop structure constants $\langle a^{(1)}_{\tau_0,\ell}\rangle$ one again needs to use the formula \eqref{eq:modified.structure.constant}.

\subsection{Complete One-Loop Resummation}
In the previous section we found the CFT data for all twists and spins. We can now supplement it into the conformal partial wave expansion \eqref{eq:BlockDecomposition} and reproduce the full one-loop correlation function.  After we do that we need to check the obtained function indeed satisfies the bootstrap equation \eqref{eq:crossing.Konishi}. We have performed this calculation explicitly and have found that the crossing relation for such obtained function implies one more constraint on the parameters of our ansatz, namely
\begin{equation}
\mu_0=-\nu_0.
\end{equation}
Implementing this constraint we end up with the function
\begin{equation}\label{eq:oneloop.linear.combination}
\mathcal G^{(1)}(u,v)=\alpha_{11}\mathcal G_{11}(u,v)+\alpha_{10}\mathcal G_{10}(u,v)+(\alpha_{00}-2\,\zeta_2\,\alpha_{11})\mathcal G_{00}(u,v)+\nu_0\mathcal G_{\nu_0}(u,v)+\gamma_{\mathrm{ext}}\mathcal G_{\mathrm{ext}}(u,v),
\end{equation}
where the individual functions are given by
\begin{align}
\mathcal G_{11}(u,v)&=c\frac{u(1+u^2+v^2-2u-2v-2uv)}v\Phi(u,v),
\\
\mathcal G_{10}(u,v)&=c\frac{u\left((1+v-2u)\log u+(1+u-2v)\log v\right)}{v},
\\
\mathcal G_{00}(u,v)&=c\frac uv(1+u+v),
\\
\mathcal G_{\nu_0}(u,v)&=-c\frac{u(u+v+uv)}v\Phi(u,v),
\\
\mathcal G_{\mathrm{ext}}(u,v)&=\left({u^2}+\frac{u^2}{v^2}+c\frac{2u^2}v \right)\log u+\left(cu-\frac{u^2}{v^2}-c\frac uv-c\frac{u^2}v\right)\log v.
\end{align}
Here we introduced the usual box function \cite{Usyukina:1992jd}
\begin{equation}
\Phi(u,v)=\frac{\log \left(\frac{1-\xx_1}{1-\xx_2}\right) \log
   \left(\xx_1\, \xx_2\right)+2\left( \mathrm{Li}_2(\xx_1)- \mathrm{Li}_2(\xx_2)\right)}{\xx_1-\xx_2}.
\end{equation}
Notice that we may interpret the contribution $\mathcal{G}_{00}(u,v)$ in \eqref{eq:oneloop.linear.combination} as a one-loop renormalisation of the constant $c$. We also emphasise that the solution $\mathcal{G}_{\nu_0}(u,v)$, which produces truncated CFT data for leading twist, does not belong to the family of truncated solutions found in \cite{Heemskerk:2009pn} since it contributes to all spins for $\tau_0>2$.

Let us now come back to a general ansatz for the truncated solution with $L>0$. We can repeat all the calculations we performed in this section and we find that the solution is even more constrained than in the $L=0$ case. Working with the general ansatz we find that there is no new solution to the bootstrap equation for higher truncated spins. Namely, we find
\begin{equation}
\mu_\ell=0,\qquad \nu_\ell=0,\qquad \mathrm{for}\,\,\ell=2,4,\ldots,L.
\end{equation}
Notice that it is a similar conclusion to the one found in \cite{Caron-Huot:2017vep}. 
 
\subsection{Comparing with Konishi}
In the previous section we have found the most general one-loop four-point correlator of four identical scalars with classical dimension $\Delta_0=2$. In this section we will find the values for all the constants which selects the Konishi solution from the family \eqref{eq:oneloop.linear.combination}. The best case scenario would be to use the properties of conformal field theories to do that. One additional piece of information which we could use is the fact that the CFT data for the stress-energy tensor, which is present in the OPE of two Konishi operators, are known. It is, however, often difficult to access this information since the stress-energy tensor is not the only operator with twist $\tau_0=2$ and spin $\ell=2$ present in the OPE of two Konishi operators. For that reason we are not able to fix the Konishi four-point correlator directly from conformal symmetry and  we will need to refer to some explicit results of direct perturbative calculations which can be found in the literature. 

In particular, we start by noticing that the Konishi operator is the only operator of twist $\tau_0=2$ and spin $\ell=0$ in the OPE of two Konishi operators. For that reason the average $\langle a^{(1)}_{2,0}\rangle=
a^{(1)}_{\mathcal{K}\mathcal{K}\mathcal{K}}:=
2\,C^{(0)}_{\mathcal{K}\mathcal{K}\mathcal{K}}C^{(1)}_{\mathcal{K}\mathcal{K}\mathcal{K}}$ is the one-loop structure constant of three Konishi operators and $\overline{\langle\gamma_{2,0}\rangle}=\gamma^{(1)}_{\mathcal{K}}$ is the one-loop anomalous dimension of Konishi operator. These can be extracted from the results in \cite{Bianchi:2001cm} and in the normalisation we use in this paper they take the values
\begin{equation}\label{eq:Konishi.res}
\gamma_{\mathrm{ext}}=\overline{\langle\gamma_{2,0}\rangle}=3,\qquad \langle a^{(1)}_{2,0}\rangle=-18c.
\end{equation}   
Moreover, the averages of leading twist anomalous dimensions for all spins can be calculated using the results from \cite{Kotikov:2001sc}, see also \cite{Anselmi:1998ms}, and they provide us with the following result
\begin{equation}\label{eq:gamma.an.res}
\overline{\langle \gamma_{2,\ell} \rangle}=2S_1(\ell),\qquad \mathrm{for} \quad\ell>0.
\end{equation}
In fact, the first two values of \eqref{eq:gamma.an.res}, together with \eqref{eq:Konishi.res}, are enough to fix all the constants and we get
\begin{equation}
\alpha_{11}=-\frac12,\quad\alpha_{10}=0,\quad \alpha_{00}=-6-\zeta_2,\quad \nu_0=3,\quad\gamma_{\mathrm{ext}}=3.
\end{equation}
Substituting this in \eqref{eq:oneloop.linear.combination} we find
\begin{align}\nonumber
\mathcal{G}^{(1)}_{\mathcal{K}\mathcal{K}\mathcal{K}\mathcal{K}}(u,v)&=-\frac{c\, u}{2v}\left(1+4u+4v+4uv+u^2+v^2\right)\Phi(u,v)-6\frac{c\, u}{v}\left(1+u+v\right)\\\label{eq:Konishi.final}
&\quad+\frac{3u}{v}\left(\frac{u}{v}+uv+2cu\right)\log u+\frac{3u}{v}\left(-\frac{u}{v}-c-cu+cv \right)\log v,
\end{align}
which exactly agrees with the result in \cite{Bianchi:2001cm}. We have therefore shown that the one-loop four-point correlation function of four Konishi operators belongs to our family of solutions, and we have found the explicit values of the constants describing this solution.

\section{The Superconformal Case}
\label{sec:super.case}
In this section we will focus on the four-point function of half-BPS operators. We  follow very closely the logic from the previous section and adapt it to the case of superconformal partial wave expansion. Following the observations in section~\ref{sec:super.four.points}, the computations in this case are very similar and here we will only highlight the differences and the results. 

The most relevant difference compared to the Konishi case is that the partial waves  take a different form, we need to replace the ordinary conformal blocks by superconformal blocks. From \eqref{eq:superconformal.decomposition} it boils down to the replacement
\begin{equation}
G_{\tau,\ell}(u,v)\to u^{-2}G_{\tau+4,\ell}(u,v).
\end{equation}
Importantly, the superconformal blocks are eigenvectors of the shifted quadratic Casimir operator of the superconformal group
\begin{equation}
\mathcal C_S( u^{-2}G_{\tau_0+4,\ell}(u,v))=\mathcal J_{\tau_0}^2u^{-2}G_{\tau_0+4,\ell}(u,v).
\end{equation}
Here we have defined
\begin{equation}\label{eq:supercasimir}
\mathcal C_S=u^{-2}\mathcal C u^2+\frac{\tau_0(\tau_0-6)}4-\frac{(\tau_0+4)(\tau_0-2)}4=u^{-2}\mathcal C u^2-2\tau_0+2,
\end{equation}
so that the eigenvalue is 
\begin{equation}
\mathcal J_{\tau_0}^2=J^2_{\tau_0+4}=\left(\frac{\tau_0}2+\ell+1\right)\left(\frac{\tau_0}2+\ell+2\right).
\end{equation}
Led by these observations we define H-functions in the supersymmetric case to be
\begin{equation}
\boldsymbol{H}_{\tau_0}^{(m,\log^n)}(u,v)=\sum_{\ell=0}\langle A^{(0)}_{\tau_0,\ell}\rangle\frac{(\log \mathcal J_{\tau_0})^n}{\mathcal J_{\tau_0}^{2m}}u^{-2}G_{\tau_0+4,\ell}(u,v),
\end{equation}
where $\langle A^{(0)}_{\tau_0,\ell}\rangle$ are the structure constants \eqref{eq:structure.constants.born}.
The H-functions satisfy again a recursion relation
\begin{equation}\label{eq:recursion.super.H}
\boldsymbol{H}_{\tau_0}^{(m,\log^n)}(u,v)=\mathcal C_S\boldsymbol{H}_{\tau_0}^{(m+1,\log^n)}(u,v).
\end{equation}
Following similar arguments to the ones presented in section~\ref{sec:twist.conformal.blocks} one can prove that the power divergent part of H-functions factorises 
\begin{equation}
\boldsymbol{H}_{\tau_0}^{(m,\log^n)}(\xx_1,\xx_2)\doteq\frac{\xx_1^{-1}}{\xx_2-\xx_1}k_{\frac{\tau_0}{2}+1}(\xx_1)\overline{\boldsymbol{H}}_{\tau_0}^{(m,\log^n)}(\xx_2),
\end{equation}
where we have again defined H-function depending only on $\xx_2$ as
\begin{equation}
\overline{\boldsymbol{H}}_{\tau_0}^{(m,\log^n)}(\xx_2)=\xx_2^{-1}\sum_{\ell=0}\langle A^{(0)}_{\tau_0,\ell}\rangle\frac{\log^n \mathcal J_{\tau_0}}{\mathcal J_{\tau_0}^{2m}}\,
k_{\frac{\tau_0}{2}+\ell+2}(\xx_2).
\end{equation}
Also, the action of Casimir operator \eqref{eq:supercasimir} simplifies when acting on the power divergent part
\begin{equation}
\mathcal C_S\boldsymbol{H}_{\tau_0}^{(m,\log^n)}(\xx_1,\xx_2)\doteq\frac{\xx_1^{-1}}{\xx_2-\xx_1}k_{\frac{\tau_0}{2}+1}(\xx_1)\overline{\mathcal D}_S\overline{\boldsymbol{H}}_{\tau_0}^{(m,\log^n)}(\xx_2),
\end{equation}
where we defined
\begin{equation}
\overline{\mathcal D}_S=\xx_2^{-2}\,\overline{\mathcal D}\, \xx_2^2=-\xx_2+(2-3\xx_2)\xx_2\partial_2+(1-\xx_2)\xx_2^2\partial_2^2\,.
\end{equation}
Finally, the H-functions $\overline{\boldsymbol{H}}^{(m,\log^n)}(\xx_2)$ satisfy the following recursion relation
\begin{equation}\label{eq:recursion.super.Hbar}
\overline{\boldsymbol{H}}_{\tau_0}^{(m,\log^n)}(\xx_2)= \overline{\mathcal D}_S\overline{\boldsymbol{H}}_{\tau_0}^{(m+1,\log^n)}(\xx_2).
\end{equation}

In the following, we will compute the one-loop perturbative correction to the function $\mathcal H(u,v)$, in exactly the same way as we did in the ordinary, non-superconformal case. In particular, in analogy with \eqref{eq:oneloop.in.Hfunctions} its power divergent part can be expanded using H-functions as
\begin{align}
\mathcal H(\xx_1,\xx_2)\doteq \sum_{\tau_0}\frac{ \xx_1^{-1}}{\xx_2-\xx_1}\sum_\rho\left( A_{\tau_0,\rho}\,k_{\frac{\tau_0}{2}+1}(\xx_1)+B_{\tau_0,\rho}\,\left(\frac\partial{\partial\tau} k_{\frac{\tau}{2}+1}(\xx_1)\right)\big|_{\tau\to\tau_0}\right) \overline {\boldsymbol{H}}_{\tau_0}^{\,\rho}(\xx_2).\label{eq:oneloop.in.Hfunctions.super}
\end{align}
Here, $A_{\tau_0,\rho}$ and $B_{\tau_0,\rho}$ are large-$\mathcal{J}$ expansion coefficients of the modified structure constants and the anomalous dimensions, respectively. Again, in order to extract the CFT data, we will need only an explicit form of the power divergent part of the H-functions for $\rho=(m,\log^n)$ with $m\leq0$ and $n=0,1$. All these functions can be easily obtained from $\overline{\boldsymbol{H}}_{2}^{(0)}(\xx_2)$ and $\overline{\boldsymbol{H}}_{2}^{(0,\log)}(\xx_2)$ using the recursion relation \eqref{eq:recursion.super.Hbar} and
\begin{equation}
\overline{\boldsymbol{H}}_{\tau_0}^{(m,\log^n)}(\xx_2)\doteq \frac{\Gamma(\frac{\tau_0}{2}+1)^2}{\Gamma(\tau_0+1)}\frac1{\tilde c}\left(\left(
\tilde c(-1)^{\frac{\tau_0}2}-\tfrac{\tau_0}{2}(\tfrac{\tau_0}{2}+1)\right)\overline{\boldsymbol{H}}_{2}^{(m,\log^n)}(\xx_2)+
\overline{\boldsymbol{H}}_{2}^{(m-1,\log^n)}(\xx_2)
\right).
\end{equation}

In the superconformal case, we have not been able to compute the exact form of the complete $\boldsymbol{H}_{2}^{(0)}(u,v)$, in contrast to the conformal case. Therefore, in principle, both $\overline{\boldsymbol{H}}_{2}^{(0)}(\xx_2)$ and $\overline{\boldsymbol{H}}_{2}^{(0,\log)}(\xx_2)$ could contain enhanced divergent terms proportional to $\log^2(1-\xx_2)$. It turns out that this is not the case\footnote{The fact that we can take $\overline{\boldsymbol{H}}_{2}^{(0)}(\xx_2)$ free from powers of logarithms can be seen by explicitly computing the power divergent terms of $\overline{\boldsymbol{H}}_{2}^{(m)}(\xx_2)$ for some $m<0$ using the kernel method, and see that they can be obtained by acting $m$ times with $\overline{\mathcal D}_S$ on $\overline{\boldsymbol{H}}_{2}^{(0)}(\xx_2)$.} and we end up with expressions analogous to the conformal case
\begin{align}
\overline{\boldsymbol{H}}_{2}^{(0)}(\xx_2)&=\frac{\tilde{c}}{1-\xx_2},\\\label{eq:H0log.super}
\overline{\boldsymbol{H}}_{2}^{(0,\log)}(\xx_2)&=-\frac{\log(1-\xx_2)}{2(1-\xx_2)}\tilde c-\frac{\gamma_E}{1-\xx_2}\tilde c+
\left(
-\frac{1}{12}-\frac{1-\xx_2}{15}+\ldots
\right)\tilde c\log^2(1-\xx_2).
\end{align}
More terms in the expansion of $\overline{\boldsymbol{H}}_{2}^{(0,\log)}(\xx_2)$ can be found in appendix~\ref{app:H0log}.

Equipped with the supersymmetric H-functions we are now ready to find the form of one-loop correction to the function $\mathcal{H}(u,v)$. Following a similar discussion as in section~\ref{sec:ansatz}, we start by observing that again all power divergent contributions to $\mathcal{H}(u,v)$ are completely captured by the twist-two operators. These terms come either from an infinite towers of twist-two operators or from solution truncated in spin. The general ansatz for leading-$u$ contribution of $\mathcal{H}(u,v)$ is therefore
\begin{align}\label{eq:Hansatz}
\mathcal{H}^{(1)}_{\mathrm{L.T.}}(u,v)&=\frac{u}{v} \left(\beta_{11}\log u \log v+\beta_{10}\log u+\beta_{01}\log v+\beta_{00}\right)\tilde c+\ldots\\&+\sum_{\ell=0}^L\langle A^{(0)}_{2,\ell}\rangle u^{-2}\left( \kappa_\ell \,G_{6,\ell}(u,v)+\lambda_\ell \left(\partial_{\tau} G_{\tau+4,\ell}(u,v)\right)\big|_{\tau\to2}  \right).
\end{align} 
for some $L$. The bootstrap equation \eqref{eq:superbootstrap} immediately implies that
\begin{equation}
\beta_{10}=\beta_{01}.
\end{equation}
Moreover, by direct application of the method described in the previous section, one can check that the truncated solutions cannot be completed to a crossing symmetric function. It implies that
\begin{equation}
\kappa_\ell=0\,,\qquad \lambda_\ell=0\,, \qquad\mathrm{for}\quad\ell=0,2,4,\ldots,L .
\end{equation} 
This stays in contrast to the ordinary conformal case where the spin-zero truncated solution was allowed.

We now use the H-function method explained in section \ref{sec:method} to complete the power divergent part of \eqref{eq:Hansatz} to a full leading-$u$ answer. In particular, the H-function method allows us to find the CFT data for twist-two operators
\begin{align}\label{eq:BPS.gamma}
\overline{\langle \gamma_{2,\ell}\rangle}&=-4\beta_{11}S_{1}(\ell+2)+2\beta_{10},\\\label{eq:BPS.aa}
\overline{\langle \hat{\alpha}_{2,\ell}\rangle}&=-2\beta_{10}S_{1}(\ell+2)+\beta_{00}.
\end{align}
We could in principle continue as in the previous section and find a general solution as a function of three constants $(\beta_{11},\beta_{10},\beta_{00})$. Instead we will focus purely on the case of four half-BPS operators for which we can use additional information about the CFT data found in the literature. In particular, it is known that the twist-two operators are not degenerate and the anomalous dimensions $\gamma_{2,\ell}$ have been found by direct calculations in e.g. \cite{Plefka:2012rd}
\begin{equation}\label{eq:BPS.an.res}
\gamma_{2,\ell}=2S_1(\ell+2)\,,\qquad \ell=0,2,4,\ldots.
\end{equation} 
Additionally, the structure constants for two half-BPS operators and twist-two operators can also be found in \cite{Plefka:2012rd} and for $\ell=0$ it is
\begin{equation}\label{eq:BPS.str.cons.res}
a^{(1)}_{2,0}=-\tilde c.
\end{equation} 
Using the first two values in \eqref{eq:BPS.an.res} together with \eqref{eq:BPS.str.cons.res} we can fix our constants to\footnote{Notice that these values could also be found by considering \eqref{eq:BPS.gamma} and \eqref{eq:BPS.aa} for $\ell=-2$. This should correspond to a BPS current in the symmetric traceless representation of R-symmetry which implies $\gamma_{2,-2}=0$ and $a^{(1)}_{2,-2}=0$. It leads to $\beta_{10}=0$ and $\beta_{00}=2\, \zeta_2\,\beta_{11}$. The remaining constant can be reabsorbed into the definition of the coupling constant, leading to the result \eqref{eq:BPS.four.point}.}
 \begin{equation}\label{eq:BPS.four.point}
 \beta_{11}=-\frac{1}{2},\qquad \beta_{10}=0,\qquad\beta_{00}=-\zeta_2 .
 \end{equation}
Then the leading-$u$ result takes the form
\begin{equation}
\mathcal{H}_{\mathrm{L.T.}}(u,v)=-\tilde c\,\frac{ \xx_1 \left(2\,\text{Li}_2(\xx_2)+\left(\log \left(\xx_1\right)+\log \left(\xx_2\right)\right)\log \left(1-\xx_2\right) \right)}{2 \left(1-\xx_2\right)}.
\end{equation}

Now we can use the bootstrap equation \eqref{eq:superbootstrap} to find the complete power divergent part of the function $\mathcal{H}(u,v)$. Subsequently, we use the H-function method to find the CFT data for all twists which we collect in appendix \ref{app:superconformal.results}. Plugging it back to the superconformal partial wave decomposition we can find the complete one-loop correlator which takes the form 
\begin{equation}\label{eq:BPS.final}
\mathcal{H}(u,v)=-\frac{\tilde c\,u}{2\,v}\Phi(u,v).
\end{equation}
This agrees with the known one-loop result for the four-point correlation function of four half-BPS operators in $\mathcal{N}=4$ SYM found in  \cite{Arutyunov:2001mh}.

\section{Conclusions and Outlook}
In this paper we found a family of solutions of to conformal bootstrap equation relevant for the one-loop perturbation of four-dimensional conformal gauge theories. We employed twist conformal blocks which allow a systematic expansion around the light-cone limit, namely $u=0$, $v=0$. Starting from the most general leading expansion \eqref{eq:G.leading.twist} we were able to complete it to a full crossing symmetric function of the cross-ratios. For four-point correlator of scalar operators with dimension $\Delta=2+g \,\gamma_{\mathrm{ext}}+\mathcal{O}(g^2)$ we found a four-parameter family of solutions. By supplementing this by a few additional pieces of CFT data for the leading-twist spectrum of the theory, we extracted the known form of one-loop correlator of four Konishi operators. Repeating this analysis for half-BPS operators $\mathcal{O}_\mathbf{20'}$ in $\mathcal{N}=4$ SYM and employing the superconformal partial wave expansion we have also found an explicit form of the one-loop correlator of four such operators. 

There are many directions one could pursue using the method we described in this paper. First of all, the four-point correlator of Konishi operators is only one representative of the family of solutions we found. A natural question is whether we can identify how other scalar correlators fit into our solution. Secondly, it should be possible to generalise our construction and apply it to correlation functions of operators with higher classical dimension. This would allow to find a large class of one-loop correlation functions in conformal gauge theories. Furthermore, there should be no conceptual obstruction to generalise it to mixed correlators.

The H-function technology can be in principle applied also to higher orders in the perturbation theory. Also in this case, the CFT data can be expanded around the infinite spin and one can extract expansion coefficients for infinite towers of operators by focusing on the enhanced divergent part of the four-point function. In contrast with the one-loop case, where the complete enhanced divergent part was captured by power divergent terms, at higher orders it is possible to get other types of enhanced divergences. For example, at two loops there can be terms proportional to $\log^2 v$ which were prohibited by the conformal partial wave expansion and bootstrap equation at one loop, see section \ref{sec:ansatz}. By examining an explicit form of conformal blocks and using the bootstrap equation it is easy to see that all such contributions come from $\langle (\gamma_{\tau_0,\ell}^{(1)})^2\rangle$. They are therefore determined by the one-loop CFT data. Unfortunately, we are unable to access this information from our previous discussion since there is a degeneracy in the spectrum. It implies that, in general, $\langle (\gamma_{\tau_0,\ell}^{(1)})^2\rangle\neq \langle (\gamma_{\tau_0,\ell}^{(1)})\rangle^2$ and therefore we cannot use the one-loop averages we have calculated to determine the enhanced divergent part of the two-loop answer. In order to find it we would need to solve the mixing problem at one loop completely. This has been successfully done for the large-$N$ expansion of the correlators of four half-BPS operators in \cite{Alday:2017xua,Aprile:2017bgs,Aprile:2017xsp}. There, it has been possible to solve the mixing problem by using the knowledge of an infinite family of one-loop four-point correlators $\langle \mathcal{O}_p(x_1)\mathcal{O}_p(x_2)\mathcal{O}_q(x_3)\mathcal{O}_q(x_4)\rangle$, for $p,q\geq 2$, where $\mathcal{O}_p(x)$ is an $\mathcal{N}=4$ SYM half-BPS operator with R-symmetry labels $[0,p,0]$. Similar analysis should be possible also at weak coupling. In particular, it would allow us to find the two-loop correlation function of four Konishi operators, which is not known at the moment. We postpone it to  future work.

\section*{Acknowledgements}
We are very grateful to Fernando Alday for many inspiring discussions and for his encouragement during the different stages of writing this paper. We would also like to thank Mark van Loon for helpful discussions. TL is supported by ERC STG grant 306260. TL is grateful to the Munich Institute for Astro- and Particle Physics (MIAPP) for their hospitality during the final stages of preparing this paper.

\appendix
\section{Appendices}
\subsection{Superconformal blocks}
\label{app:superblocks}
In this appendix we present an explicit form of the superconformal blocks appearing in the expansion of correlation functions of four half-BPS operators in $\mathcal{N}=4$ SYM. We closely follow \cite{Doobary:2015gia} and restrict to the case $p_1=p_2=p_3=p_4=2$, which is the one relevant for this paper. All supermultiplets appearing in the intermediate channel of such correlation functions can be labelled by a Young tableaux $\underline\lambda=[\lambda_1,\lambda_2]$, with $\lambda_1\geq\lambda_2$, consisting of maximally two rows, and a charge $\gamma=0,2,4$. We distinguish three types of multiplets: half-BPS, quarter-BPS and long, whose representation labels are summarised in the table~\ref{tab:supermultiplets}. Notice that the only long multiplets are in the singlet representation $[0,0,0]$ of the $SU(4)$ R-symmetry. 

\begin{table}[ht]
\centering
\begin{tabular}{|c|c|c|c|c|}
\hline
Young tableaux $\underline\lambda$& twist $\tau$ & spin $\ell$&R-symmetry representation& multiplet type\\
\hline
$[0,0]$&$\gamma$&0&$[0,\gamma,0]$&half-BPS\\
\hline
$[\lambda_1,0],\lambda_1\geq 2$&$\gamma$&$\lambda_1-2$&$[0,\gamma-2,0]$&quarter-BPS\\
$[\lambda_1,1],\lambda_1\geq 2$&$\gamma$&$\lambda_1-2$&$[1,\gamma-4,1]$&quarter-BPS\\
$[1,0]$&$\gamma$&$0$&$[1,\gamma-2,1]$&quarter-BPS\\
$[1,1]$&$\gamma$&$0$&$[2,\gamma-4,2]$&quarter-BPS\\
\hline
$[\lambda_1,\lambda_2],\lambda_2\geq 2$&$2\lambda_2$&$\lambda_1-\lambda_2$&$[0,0,0]$&long\\
\hline
\end{tabular}
\caption{Supermultiplets appearing in the superconformal partial waves of $\langle \mathcal{O}_{\mathbf{20'}}\mathcal{O}_{\mathbf{20'}}\mathcal{O}_{\mathbf{20'}}\mathcal{O}_{\mathbf{20'}}\rangle$.}\label{tab:supermultiplets}
\end{table}

The superconformal blocks are given by
\begin{equation}
\mathcal{S}_{\mathcal{R}}(x,y)=\left(\frac{\xx_1 \xx_2}{\yy_1 \yy_2}\right)^{\gamma/2}\mathcal{F}^{\gamma, \underline \lambda}(\xx,\yy),
\end{equation}
where
\begin{equation}
\mathcal{F}^{\gamma, \underline \lambda}(\xx,\yy)=(-1)^{\frac{\gamma}{2}-1}D^{-1}\,\det \begin{pmatrix}
F^X_{\underline \lambda}(\xx)&R\\K_{\underline\lambda}&F^Y(\yy)
\end{pmatrix}.
\end{equation}
The explicit form of all ingredients (with $1\leq i,j\leq 2$ and $1\leq m,n\leq\gamma/2$) is
\begin{align}
(F_{\underline\lambda}^X(\xx))_{in}&=[\xx_i^{\lambda_n-n}{_2F_1}(\lambda_n+1-n+\tfrac{\gamma}{2},\lambda_n+1-n+\tfrac{\gamma}{2},2\lambda_n+2-2n+\gamma;\xx_i)],\\
(F^Y(y))_{mj}&=(y_j)^{m-1}{}_2F_1(m-\tfrac{\gamma}{2},m-\tfrac{\gamma}{2},2m-\gamma;y_j),\\
(K_{\underline\lambda})_{mn}&=-\delta_{m,n-\lambda_n},\\
R&=\begin{pmatrix}
\frac{1}{\xx_1-\yy_1}&\frac{1}{\xx_1-\yy_2}\\\frac{1}{\xx_2-\yy_1}&\frac{1}{\xx_2-\yy_2}
\end{pmatrix},\\
D&=\frac{(\xx_1-\xx_2)(\yy_1-\yy_2)}{(\xx_1-\yy_1)(\xx_1-\yy_2)(\xx_2-\yy_1)(\xx_2-\yy_2)}.
\end{align}  
 Here, the square bracket in the definition of $F^X$ indicates that we keep only the regular part, namely
 \begin{equation}
 [\xx^{-\alpha}{}_2F_1(a,b,c;\xx)]=\xx^{-\alpha}{}_2F_1(a,b,c;\xx)-\sum_{k=0}^{\alpha-1}\frac{(a)_k(b)_k}{(c)_k k!}\xx^{k-\alpha}=\sum_{k=0}^{\infty}\frac{(a)_{k+\alpha}(b)_{k+\alpha}}{(c)_{k+\alpha}(k+\alpha)!}\xx^{k}\,.
 \end{equation}
Importantly, for long multiplets have $\gamma=4$, $\lambda_2=\tfrac{\tau}{2}$, $\lambda_1=\ell+\tfrac{\tau}{2}$, $\tau\geq4$ and $\alpha\geq 0$. Then, the superconformal blocks can be written in a more explicit form as
\begin{align}\label{eq:super.long}
\mathcal{F}_{\mathrm{long}}(\xx,\yy)=\frac{(\xx_1-\yy_1)(\xx_1-\yy_2)(\xx_2-\yy_1)(\xx_2-\yy_2)}{(\xx_1 \,\xx_2)^4}G_{\tau+4,\ell}(\xx_1,\xx_2) \,,
\end{align}   
where $G_{\tau,\ell}(\xx_1 ,\xx_2)$ is the ordinary conformal block in four dimensions \eqref{eq:ConformalBlock} as found in~\cite{Dolan:2001tt}.
  
At the unitarity bound, quarter-BPS multiplets can combine to form a long multiplet in the interacting theory. This is exactly the case for the twist-two multiplets in the singlet representation
\begin{equation}
(\gamma=2,\underline\lambda=[\ell+2,0])\oplus (\gamma=4,\underline\lambda=[\ell+1,1])\longrightarrow (\gamma=4,\underline\lambda=[\ell+1,1])_\mathrm{long}\,.
\end{equation}
Using the explicit form of superconformal blocks one can write
\begin{equation}\label{eq:twist2.recomb}
\frac{\yy_1 \yy_2}{\xx_1 \xx_2}\mathcal{F}^{2,[\ell+2,0]}(x,y)+\mathcal{F}^{4,[\ell+1,1]}(x,y)=\frac{(\xx_1-\yy_1)(\xx_1-\yy_2)(\xx_2-\yy_1)(\xx_2-\yy_2)}{(\xx_1 \,\xx_2)^4}G_{6,\ell}(\xx_1,\xx_2) ,
\end{equation} 
which agrees with \eqref{eq:super.long} for $\tau=2$.

\subsection{More Details On $\overline{H}^{(0,\log)}(\xx_2)$}
\label{app:H0log}
In the expression~\eqref{eq:H0log} for $\overline H^{(0,\log)}(\xx_2)$, the coefficients $e_i$ multiplying $(1-\xx_2)^i\log^2(1-\xx_2)$ for $i=\{0,1,2,\ldots\}$ are given by the sequence
\begin{equation}
\left\{-\frac{1}{12},\frac{1}{10},-\frac{5}{504},-\frac{8}{2835} ,-\frac{251}{199584}, 
-\frac{55967}{81081000},-\frac{2499683}{5837832000},-\frac{50019793}{173675502000},
\ldots 
\right\}.
\end{equation}
The corresponding values in the superconformal case \eqref{eq:H0log.super} are
\begin{equation}
\left\{-\frac{1}{12},-\frac{1}{15},-\frac{151}{2520},-\frac{127}{2268},-\frac{53219}{997920},-\frac{8327609}{162162000},-\frac{290756381}{5837832000},-\frac{5620770149}{115783668000},
\ldots
\right\}.
\end{equation}

\subsection{Modified Structure Constants for the Conformal Case}
\label{app:results.nonsuper}
In this appendix we present an exact form of the modified structure constants that appear in \eqref{eq:structure.constants.sum}. The equations below are valid for $\tau_0>2$.
\begin{align}
\overline{\langle\hat\alpha_{\tau_0,\ell}\rangle}_{11}&=\frac{4\,c\,\eta}{P_{\tau_0,\ell}}\Big(
-\zeta_2+S_1\left( \tfrac{\tau_0}{2}-2 \right)^2-S_1\left( \tfrac{\tau_0}{2}-2 \right)S_1\left( \tau_0-4 \right)-\frac12 S_2\left( \tfrac{\tau_0}{2}-2 \right)-\tfrac{\delta_{\tau_0,4}}{2}
\nonumber\\
&\quad \qquad + \left[2S_1\left( \tfrac{\tau_0}{2}-2 \right)-S_1\left(\tau_0-4\right)+\tfrac{\delta_{\tau_0,4}}{4}\right]S_1\left( \tfrac{\tau_0}{2}+\ell-1 \right)
\Big),
\\
\overline{\langle\hat\alpha_{\tau_0,\ell}\rangle}_{10}&=\frac{2\,c\,\eta}{P_{\tau_0,\ell}}\Big(-3S_1\left( \tfrac{\tau_0}{2}-2 \right)+2S_1\left(\tau_0-4\right)-\tfrac{3\delta_{\tau_0,4}}{4}-S_1\left( \tfrac{\tau_0}{2}+\ell-1 \right)\Big),
\\
\overline{\langle\hat\alpha_{\tau_0,\ell}\rangle}_{00}&=\frac{c\,\eta}{P_{\tau_0,\ell}},
\\
\overline{\langle\hat\alpha_{\tau_0,\ell}\rangle}_{\mathrm{ext}}&=\frac{2\,c\,\eta}{P_{\tau_0,\ell}}\Big(1+ S_1\left( \tfrac{\tau_0}{2}-2 \right)-S_1\left( \tau_0-4 \right)+\frac{\delta_{\tau_0,4}}{2}\Big)-\frac{\tau_0-3}{P_{\tau_0,\ell}}
\nonumber\\
&\quad+2\left[-1+2 S_1\left( \tfrac{\tau_0}{2}-2 \right)-S_1\left(\tau_0-4 \right)+S_1\left( \tfrac{\tau_0}{2}+\ell-1 \right) \right],
\\
\overline{\langle\hat\alpha_{\tau_0,\ell}\rangle}_{\mu_0}&=\frac{4\,c}{P_{\tau_0,\ell}}\Big(-S_1\left( \tfrac{\tau_0}{2}-2 \right)+ S_1\left(\tau_0-4\right)\Big),
\\
\overline{\langle\hat\alpha_{\tau_0,\ell}\rangle}_{\nu_0}&=\frac{2\,c}{P_{\tau_0,\ell}}\Big(-\zeta_2-2S_1\left( \tfrac{\tau_0}{2}-2 \right)-2S_1\left( \tfrac{\tau_0}{2}-2 \right)^2+2S_1\left(\tau_0-4\right) +S_2\left( \tfrac{\tau_0}{2}-2 \right)
\nonumber\\
&\quad \qquad+2S_1\left( \tfrac{\tau_0}{2}-2 \right)S_1\left(\tau_0-4\right)+2\left[S_1\left( \tfrac{\tau_0}{2}-2 \right)- S_1\left(\tau_0-4\right) \right]S_1\left( \tfrac{\tau_0}{2}+\ell-1 \right)\Big).
\end{align}
As in \eqref{eq:gamma.HT}, we have $\eta=(-1)^{\frac{\tau_0}{2}}$ and $P_{\tau_0,\ell}=c\eta+\left(\tau_0+\ell-2\right)\left(\ell+1\right)$.

\subsection{Konishi CFT Data}
We present here an explicit form of the CFT data for operators present in the conformal partial wave decomposition of \eqref{eq:Konishi.final}.

The anomalous dimensions are given by
\begin{align}
\overline{\langle \gamma_{2,\ell}\rangle}&=2S_1(\ell)+3\delta_{l,0},
\\
\overline{\langle \gamma_{\tau_0,\ell}\rangle}&=6+\frac{12 c}{P_{\tau_0,\ell}}\left[-S_1\left(\tfrac{\tau_0}{2}-2\right)+S_1\left(\tfrac{\tau_0}{2}+\ell-1\right)\right]
\nonumber\\&\quad
+\frac{c\,\eta}{P_{\tau_0,\ell}}\left[
6-\delta_{\tau_0,4}-2S_1\left(\tfrac{\tau_0}{2}-2\right)-2\left(\tfrac{\tau_0}{2}+\ell-1\right)
\right],\qquad \tau_0>2,
\end{align}
and the modified structure constants are
\begin{align}
\overline{\langle \hat\alpha_{2,\ell}\rangle}&=-6-3\delta_{\ell,0}-\zeta_2+6 S_1(\ell),
\\
\overline{\langle \hat\alpha_{\tau_0,\ell}\rangle}&=6\left[
-1+2S_1\left(\tfrac{\tau_0}{2}-2\right)-S_1\left(\tau_0-4\right)+S_1\left(\tfrac{\tau_0}{2}+\ell-1\right)
\right]-\frac{3}{P_{\tau_0,\ell}}(\tau_0-3)
\nonumber\\
&\quad+\frac{c}{P_{\tau_0,\ell}}\Big(12\left[ 
S_1\left(\tfrac{\tau_0}{2}-2\right)-S_1\left(\tau_0-4\right)
\right]
\nonumber\\&\qquad\qquad\quad
+\eta\left[-4 S_1\left(\tfrac{\tau_0}{2}-2\right)+2S_1\left(\tau_0-4\right)-\tfrac{\delta_{\tau_0,4}}{2}\right]\Big)S_1\left(\tfrac{\tau_0}{2}+\ell-1\right)
\nonumber\\
&\quad+\frac{6\,c}{P_{\tau_0,\ell}}\Big(
-\zeta_2-2S_1\left(\tfrac{\tau_0}{2}-2\right)^2+2 S_1\left(\tfrac{\tau_0}{2}-2\right)S_1\left(\tau_0-4\right)+S_2\left(\tfrac{\tau_0}2-2\right)
\Big)
\nonumber\\
&\quad+\frac{c\,\eta}{P_{\tau_0,\ell}}\Big(\zeta_2+6 S_1\left(\tfrac{\tau_0}{2}-2\right)-2S_1\left(\tfrac{\tau_0}{2}-2\right)^2-6 S_1\left(\tau_0-4\right)
\nonumber\\&\qquad\qquad\quad
+2S_1\left(\tfrac{\tau_0}{2}-2\right)S_1\left(\tau_0-4\right)+S_2\left(\tfrac{\tau_0}{2}-2\right)+4\delta_{\tau_0,4}
\Big) , \qquad \tau_0>2.
\end{align}
Recall that the one-loop structure constants $\langle a^{(1)}_{\tau_0,\ell}\rangle$ can be found using \eqref{eq:modified.structure.constant}.

\subsection{Half-BPS CFT Data}
\label{app:superconformal.results}
We present here an explicit form of the CFT data for long supermultiplets present in the superconformal partial wave decomposition of \eqref{eq:BPS.final}.

The anomalous dimensions are given by
\begin{align}
\overline{\langle \gamma_{2,\ell}\rangle}&=2S_1(\ell+2),
\\
\overline{\langle \gamma_{\tau_0,\ell}\rangle}&=-\frac{2\,\tilde c}{\mathcal P_{\tau_0,\ell}}\Big((\eta+1)S_1\left(\tfrac{\tau_0}2\right)+(\eta-1)S_1\left(\tfrac{\tau_0}2+\ell+1\right) \Big), \qquad \tau_0>2\,,
\end{align}
and the modified structure constants are
\begin{align}
\overline{\langle \hat\alpha_{2,\ell}\rangle}&=-\zeta_2,
\\
\overline{\langle \hat\alpha_{\tau_0,\ell}\rangle}&=-\frac{2\,\tilde c}{\mathcal P_{\tau_0,\ell}}\Big(\left[
(2\eta-1)S_1\left(\tfrac{\tau_0}2\right)+(1-\eta)S_1(\tau_0)
\right]S_1\left(\tfrac{\tau_0}2+\ell+1\right)+(1+\eta)S_1\left(\tfrac{\tau_0}{2}\right)^2
\nonumber\\
&\qquad\qquad
-(1+\eta)S_1\left(\tfrac{\tau_0}{2}\right)S_1(\tau_0)-\frac{1+\eta}{2}S_2\left(\tfrac{\tau_0}{2}\right)+\frac{1-\eta}{2}\zeta_2
\Big),\qquad \tau_0>2,
\end{align}
where $\mathcal P_{\tau_0,\ell}=\tilde c\,\eta+(\tau_0+\ell+2)(\ell+1)$ is the factor appearing in the higher twist structure constants \eqref{eq:structure.constants.born}. The one-loop structure constants $\langle a^{(1)}_{\tau_0,\ell}\rangle$ can be found using the supersymmetric version of \eqref{eq:modified.structure.constant}.

\bibliographystyle{nb}
\bibliography{Analytic_bootstrap}

\begin{thebibliography}{10}
\ifx\href\asklfhas\newcommand{\href}[2]{#2}\fi
\ifx\arxivref\asklfhas\newcommand{\arxivref}[2]{\href{http://arxiv.org/abs/#1}{#2}}\fi
\ifx\doiref\asklfhas\newcommand{\doiref}[2]{\href{http://dx.doi.org/#1}{#2}}\fi
\raggedright
\small
\parskip 0pt

%%CITATION = ARXIV:0807.0004;%%
\bibitem{Rattazzi:2008pe}
R.~Rattazzi, V.~S.~Rychkov, E.~Tonni and A.~Vichi,
\textit{``{Bounding scalar operator dimensions in 4D CFT}''},
\textsf{\doiref{10.1088/1126-6708/2008/12/031}{JHEP~0812,~031~(2008)}},
\texttt{\arxivref{0807.0004}{arxiv:0807.0004}}.

%%CITATION = ARXIV:1203.6064;%%
\bibitem{ElShowk:2012ht}
S.~El-Showk, M.~F.~Paulos, D.~Poland, S.~Rychkov, D.~Simmons-Duffin and
  A.~Vichi,
\textit{``{Solving the 3D Ising Model with the Conformal Bootstrap}''},
\textsf{\doiref{10.1103/PhysRevD.86.025022}{Phys.~Rev.~D86,~025022~(2012)}},
\texttt{\arxivref{1203.6064}{arxiv:1203.6064}}.

%%CITATION = ARXIV:1212.4103;%%
\bibitem{Komargodski:2012ek}
Z.~Komargodski and A.~Zhiboedov,
\textit{``{Convexity and Liberation at Large Spin}''},
\textsf{\doiref{10.1007/JHEP11(2013)140}{JHEP~1311,~140~(2013)}},
\texttt{\arxivref{1212.4103}{arxiv:1212.4103}}.

%%CITATION = ARXIV:1212.3616;%%
\bibitem{Fitzpatrick:2012yx}
A.~L.~Fitzpatrick, J.~Kaplan, D.~Poland and D.~Simmons-Duffin,
\textit{``{The Analytic Bootstrap and AdS Superhorizon Locality}''},
\textsf{\doiref{10.1007/JHEP12(2013)004}{JHEP~1312,~004~(2013)}},
\texttt{\arxivref{1212.3616}{arxiv:1212.3616}}.

%%CITATION = ARXIV:1305.4604;%%
\bibitem{Alday:2013cwa}
L.~F.~Alday and A.~Bissi,
\textit{``{Higher-spin correlators}''},
\textsf{\doiref{10.1007/JHEP10(2013)202}{JHEP~1310,~202~(2013)}},
\texttt{\arxivref{1305.4604}{arxiv:1305.4604}}.

%%CITATION = ARXIV:1506.04659;%%
\bibitem{Alday:2015ota}
L.~F.~Alday and A.~Zhiboedov,
\textit{``{Conformal Bootstrap With Slightly Broken Higher Spin Symmetry}''},
\textsf{\doiref{10.1007/JHEP06(2016)091}{JHEP~1606,~091~(2016)}},
\texttt{\arxivref{1506.04659}{arxiv:1506.04659}}.

%%CITATION = ARXIV:1611.01500;%%
\bibitem{Alday:2016njk}
L.~F.~Alday,
\textit{``{Large Spin Perturbation Theory for Conformal Field Theories}''},
\textsf{\doiref{10.1103/PhysRevLett.119.111601}{Phys.~Rev.~Lett.~119,~111601~(2017)}},
\texttt{\arxivref{1611.01500}{arxiv:1611.01500}}.

%%CITATION = ARXIV:1612.00696;%%
\bibitem{Alday:2016jfr}
L.~F.~Alday,
\textit{``{Solving CFTs with Weakly Broken Higher Spin Symmetry}''},
\texttt{\arxivref{1612.00696}{arxiv:1612.00696}}.

%%CITATION = ARXIV:1706.02388;%%
\bibitem{Alday:2017xua}
L.~F.~Alday and A.~Bissi,
\textit{``{Loop Corrections to Supergravity on $AdS_5 \times S^5$}''},
\texttt{\arxivref{1706.02388}{arxiv:1706.02388}}.

%%CITATION = ARXIV:0907.0151;%%
\bibitem{Heemskerk:2009pn}
I.~Heemskerk, J.~Penedones, J.~Polchinski and J.~Sully,
\textit{``{Holography from Conformal Field Theory}''},
\textsf{\doiref{10.1088/1126-6708/2009/10/079}{JHEP~0910,~079~(2009)}},
\texttt{\arxivref{0907.0151}{arxiv:0907.0151}}.

%%CITATION = ARXIV:1410.4717;%%
\bibitem{Alday:2014tsa}
L.~F.~Alday, A.~Bissi and T.~Lukowski,
\textit{``{Lessons from crossing symmetry at large N}''},
\textsf{\doiref{10.1007/JHEP06(2015)074}{JHEP~1506,~074~(2015)}},
\texttt{\arxivref{1410.4717}{arxiv:1410.4717}}.

%%CITATION = HEP-TH/0010137;%%
\bibitem{Arutyunov:2000im}
G.~Arutyunov, S.~Frolov and A.~Petkou,
\textit{``{Perturbative and instanton corrections to the OPE of CPOs in
  $\mathcal N=4$ SYM$_4$}''},
\textsf{\doiref{10.1016/S0550-3213(01)00118-3,
  10.1016/S0550-3213(01)00265-6}{Nucl.~Phys.~B602,~238~(2001)}},
\texttt{\arxivref{hep-th/0010137}{hep-th/0010137}},
[Erratum: Nucl. Phys.B609,540(2001)].

%%CITATION = HEP-TH/0011040;%%
\bibitem{Dolan:2000ut}
F.~A.~Dolan and H.~Osborn,
\textit{``{Conformal four point functions and the operator product
  expansion}''},
\textsf{\doiref{10.1016/S0550-3213(01)00013-X}{Nucl.~Phys.~B599,~459~(2001)}},
\texttt{\arxivref{hep-th/0011040}{hep-th/0011040}}.

%%CITATION = HEP-TH/0412335;%%
\bibitem{Dolan:2004iy}
F.~A.~Dolan and H.~Osborn,
\textit{``{Conformal partial wave expansions for N=4 chiral four point
  functions}''},
\textsf{\doiref{10.1016/j.aop.2005.07.005}{Annals~Phys.~321,~581~(2006)}},
\texttt{\arxivref{hep-th/0412335}{hep-th/0412335}}.

%%CITATION = HEP-TH/0407060;%%
\bibitem{Nirschl:2004pa}
M.~Nirschl and H.~Osborn,
\textit{``{Superconformal Ward identities and their solution}''},
\textsf{\doiref{10.1016/j.nuclphysb.2005.01.013}{Nucl.~Phys.~B711,~409~(2005)}},
\texttt{\arxivref{hep-th/0407060}{hep-th/0407060}}.

%%CITATION = ARXIV:1508.03611;%%
\bibitem{Doobary:2015gia}
R.~Doobary and P.~Heslop,
\textit{``{Superconformal partial waves in Grassmannian field theories}''},
\textsf{\doiref{10.1007/JHEP12(2015)159}{JHEP~1512,~159~(2015)}},
\texttt{\arxivref{1508.03611}{arxiv:1508.03611}}.

%%CITATION = HEP-TH/0209056;%%
\bibitem{Dolan:2002zh}
F.~A.~Dolan and H.~Osborn,
\textit{``{On short and semi-short representations for four-dimensional
  superconformal symmetry}''},
\textsf{\doiref{10.1016/S0003-4916(03)00074-5}{Annals~Phys.~307,~41~(2003)}},
\texttt{\arxivref{hep-th/0209056}{hep-th/0209056}}.

%%CITATION = HEP-TH/0307210;%%
\bibitem{Heslop:2003xu}
P.~J.~Heslop and P.~S.~Howe,
\textit{``{Aspects of N=4 SYM}''},
\textsf{\doiref{10.1088/1126-6708/2004/01/058}{JHEP~0401,~058~(2004)}},
\texttt{\arxivref{hep-th/0307210}{hep-th/0307210}}.

%%CITATION = ARXIV:1502.07707;%%
\bibitem{Alday:2015eya}
L.~F.~Alday, A.~Bissi and T.~Lukowski,
\textit{``{Large spin systematics in CFT}''},
\textsf{\doiref{10.1007/JHEP11(2015)101}{JHEP~1511,~101~(2015)}},
\texttt{\arxivref{1502.07707}{arxiv:1502.07707}}.

%%CITATION = ARXIV:0708.0672;%%
\bibitem{Alday:2007mf}
L.~F.~Alday and J.~M.~Maldacena,
\textit{``{Comments on operators with large spin}''},
\textsf{\doiref{10.1088/1126-6708/2007/11/019}{JHEP~0711,~019~(2007)}},
\texttt{\arxivref{0708.0672}{arxiv:0708.0672}}.

%%CITATION = ARXIV:1510.08091;%%
\bibitem{Alday:2015ewa}
L.~F.~Alday and A.~Zhiboedov,
\textit{``{An Algebraic Approach to the Analytic Bootstrap}''},
\textsf{\doiref{10.1007/JHEP04(2017)157}{JHEP~1704,~157~(2017)}},
\texttt{\arxivref{1510.08091}{arxiv:1510.08091}}.

%%CITATION = PHLTA,B298,363;%%
\bibitem{Usyukina:1992jd}
N.~I.~Usyukina and A.~I.~Davydychev,
\textit{``{An Approach to the evaluation of three and four point ladder
  diagrams}''},
\textsf{\doiref{10.1016/0370-2693(93)91834-A}{Phys.~Lett.~B298,~363~(1993)}}.

%%CITATION = ARXIV:1703.00278;%%
\bibitem{Caron-Huot:2017vep}
S.~Caron-Huot,
\textit{``{Analyticity in Spin in Conformal Theories}''},
\textsf{\doiref{10.1007/JHEP09(2017)078}{JHEP~1709,~078~(2017)}},
\texttt{\arxivref{1703.00278}{arxiv:1703.00278}}.

%%CITATION = HEP-TH/0104016;%%
\bibitem{Bianchi:2001cm}
M.~Bianchi, S.~Kovacs, G.~Rossi and Y.~S.~Stanev,
\textit{``{Properties of the Konishi multiplet in N=4 SYM theory}''},
\textsf{\doiref{10.1088/1126-6708/2001/05/042}{JHEP~0105,~042~(2001)}},
\texttt{\arxivref{hep-th/0104016}{hep-th/0104016}}.

%%CITATION = HEP-PH/0112346;%%
\bibitem{Kotikov:2001sc}
A.~V.~Kotikov and L.~N.~Lipatov,
\textit{``{DGLAP and BFKL evolution equations in the N=4 supersymmetric gauge
  theory}''},
\texttt{\arxivref{hep-ph/0112346}{hep-ph/0112346}},
in: \textit{``{35th Annual Winter School on Nuclear and Particle Physics
  Repino, Russia, February 19-25, 2001}''}.

%%CITATION = HEP-TH/9809192;%%
\bibitem{Anselmi:1998ms}
D.~Anselmi,
\textit{``{The N=4 quantum conformal algebra}''},
\textsf{\doiref{10.1016/S0550-3213(98)00848-7}{Nucl.~Phys.~B541,~369~(1999)}},
\texttt{\arxivref{hep-th/9809192}{hep-th/9809192}}.

%%CITATION = ARXIV:1207.4784;%%
\bibitem{Plefka:2012rd}
J.~Plefka and K.~Wiegandt,
\textit{``{Three-Point Functions of Twist-Two Operators in N=4 SYM at One
  Loop}''},
\textsf{\doiref{10.1007/JHEP10(2012)177}{JHEP~1210,~177~(2012)}},
\texttt{\arxivref{1207.4784}{arxiv:1207.4784}}.

%%CITATION = HEP-TH/0103230;%%
\bibitem{Arutyunov:2001mh}
G.~Arutyunov, B.~Eden, A.~C.~Petkou and E.~Sokatchev,
\textit{``{Exceptional nonrenormalization properties and OPE analysis of chiral
  four point functions in N=4 SYM$_4$}''},
\textsf{\doiref{10.1016/S0550-3213(01)00569-7}{Nucl.~Phys.~B620,~380~(2002)}},
\texttt{\arxivref{hep-th/0103230}{hep-th/0103230}}.

%%CITATION = ARXIV:1706.02822;%%
\bibitem{Aprile:2017bgs}
F.~Aprile, J.~M.~Drummond, P.~Heslop and H.~Paul,
\textit{``{Quantum Gravity from Conformal Field Theory}''},
\texttt{\arxivref{1706.02822}{arxiv:1706.02822}}.

%%CITATION = ARXIV:1706.08456;%%
\bibitem{Aprile:2017xsp}
F.~Aprile, J.~M.~Drummond, P.~Heslop and H.~Paul,
\textit{``{Unmixing Supergravity}''},
\texttt{\arxivref{1706.08456}{arxiv:1706.08456}}.

%%CITATION = HEP-TH/0112251;%%
\bibitem{Dolan:2001tt}
F.~A.~Dolan and H.~Osborn,
\textit{``{Superconformal symmetry, correlation functions and the operator
  product expansion}''},
\textsf{\doiref{10.1016/S0550-3213(02)00096-2}{Nucl.~Phys.~B629,~3~(2002)}},
\texttt{\arxivref{hep-th/0112251}{hep-th/0112251}}.

\end{thebibliography}

\end{document}